\documentclass[fleqn]{article}
\usepackage{graphicx} 
\usepackage{subfig}
\usepackage{geometry}
\usepackage[utf8]{inputenc}
\usepackage[T1]{fontenc}
\usepackage{setspace}
\usepackage{amsmath}
\usepackage{bbm}
\usepackage{float}
\usepackage{microtype}
\usepackage{amssymb}
\usepackage{amsthm}
\newtheorem{mydef}{Definition}
\usepackage{tikz}
\usepackage{mathtools}
\usepackage{enumerate}
\usepackage{authblk}
\usepackage{hyperref}
\usepackage{xcolor}
\usepackage[
    backend=biber,
    style=nature]{biblatex} 
\newcommand{\Okl}{\Omega_{k,l}}

\newcommand{\E}{\mathcal{E}}

\newcommand{\mL}{\mathcal{L}}
\newcommand{\mP}{\mathcal{P}}
\newcommand{\mQ}{\mathcal{Q}}

\newcommand{\Zd}{\mathcal{Z}_d}
\newcommand{\Z}{\mathcal{Z}}
\newcommand{\w}{\omega}
\newcommand{\denop}{\mathcal{D}}
\newcommand{\hs}{\mathcal{H}}
\newcommand{\otAB}{\overset{AB}{\otimes}}

\newcommand{\id}{\mathbbm{1}}
\newcommand{\Oe}{| \Omega(e) \rangle}

\newcommand{\Prob}{\mathbbm{P}}

\newcommand{\Pnull}{| \Omega_{0, 0} \rangle \langle \Omega_{0,0}| }

\DeclareMathOperator{\Tr}{Tr}
\newtheorem{theorem}{Theorem}

\newtheorem{lemma}[theorem]{Lemma}
\newtheorem{prop}[theorem]{Proposition}

\title{Resource State Distillation via Stabilizer Channels}
\author[1]{Christopher Popp}
\author[2]{Tobias C. Sutter}
\author[3]{Beatrix C. Hiesmayr}
\affil[1,2,3]{University of Vienna, Faculty of Physics, Währingerstrasse 17, 1090 Vienna.\vspace{3.5mm}}
\affil[1]{christopher.popp@univie.ac.at}
\affil[2]{tobias.christoph.sutter@univie.ac.at}
\affil[3]{beatrix.hiesmayr@univie.ac.at}

\date{}

\begin{document}

\onehalfspacing
\maketitle
\begin{abstract}
Quantum technologies rely on high-quality resource states, such as maximally entangled or private states, which are indispensable for quantum communication and cryptography. In practice, however, these states are inevitably degraded by noise. Distillation protocols aim to recover high-resource states from multiple imperfect copies, and while stabilizer-based methods have demonstrated high performance in entanglement purification, they have yet to be established for broader tasks such as secret-key distillation.
This work introduces a unified framework for stabilizer-based resource distillation in systems of prime local dimension. By formulating stabilizer routines as quantum channels and deriving closed-form expressions for their output, we enable the application of stabilizer operations to general input states and diverse distillation objectives. We identify key invariances in resource measures—such as coherent and private information—and demonstrate how they can be leveraged to significantly reduce the numerical complexity of channel optimization.
To illustrate the framework's versatility, we introduce several protocols: gF-IMAX for general fidelity optimization, and (S)CI-IMAX and (S)PI-IMAX for optimizing (smooth) coherent and private information in both asymptotic and one-shot regimes. Our numerical results confirm that these protocols effectively tailor stabilizer channels to specific operational tasks, establishing them as a robust and flexible tool for quantum resource distillation.
\end{abstract}
\section{Introduction}
Quantum information processing relies fundamentally on the availability of suitable quantum resources \cite{chitambar_quantum_2019}. Depending on the task, these resources may take the form of static objects such as quantum states or dynamic objects such as quantum channels \cite{khatri_principles_2024}. Static resources include entangled states \cite{horodecki_quantum_2009},  private or secret-key states \cite{horodecki_general_2009}, and more exotic forms such as data-hiding \cite{divincenzo_quantum_2001} or locking \cite{divincenzo_locking_2004} states. Each of these enables specific operational advantages, from quantum communication to cryptographic security (cf. Ref.\cite{khatri_principles_2024}). Realistic quantum systems are inevitably affected by noise and imperfections. The process of transforming many noisy states into fewer, highly resourceful ones is called distillation and plays a central role in quantum communication and cryptography \cite{wilde_quantum_2013}.\\
Two prominent examples are entanglement distillation (also named entanglement purification) \cite{bennett_purification_1996}, which aims to produce maximally entangled states, and secret-key distillation \cite{bennett_quantum_2014-1, ekert_quantum_1991}, which generates states from which a secure classical key can be extracted. Both tasks can be studied in the asymptotic regime, where arbitrarily many copies of the input state are available, and in the resource-limited one-shot regime \cite{tomamichel_quantum_2016}, where only a single use of the distillation channel is permitted. Distillation protocols are implemented through local operations and classical communication (LOCC) in the bipartite setting, or local operations and public communication (LOPC) in the tripartite setting, including a potentially adversarial party. Their performance can be characterized by information-theoretic quantities that measure a state’s usefulness for a given task (see Ref.\cite{khatri_principles_2024} for a comprehensive summary). Examples include the coherent information, the mutual and private information \cite{wilde_quantum_2013}, and their ``smoothed'' one-shot counterparts, such as the smooth conditional max-entropy \cite{renner_security_2006} and the hypothesis-testing mutual information \cite{wang_one-shot_2012}. These quantities provide lower bounds on the distillable entanglement or secret-key and therefore serve as natural optimization targets \cite{devetak_distillation_2005, khatri_second-order_2021, wilde_converse_2017}. 
Many of these quantities can be described in terms of generalized divergences and have properties \cite{popp_local_2025} that make their computation tractable via semidefinite programming \cite{skrzypczyk_semidefinite_2023, tavakoli_semidefinite_2024, nuradha_fidelity-based_2024, popp_computation_2026}.
\\
Among the many entanglement distillation protocols \cite{bennett_purification_1996, deutsch_quantum_1996, horodecki_reduction_1998, alber_efficient_2001, miguel-ramiro_efficient_2018, dehaene_local_2003, vollbrecht_efficient_2003}, the application of stabilizer codes \cite{rains_nonbinary_1999, ashikhmin_nonbinary_2001} to  distillation \cite{matsumoto_conversion_2003} has recently proven particularly powerful \cite{popp_novel_2025, popp_low-fidelity_2025}. Leveraging a shared group structure of stabilizers and Bell states, an entanglement distillation protocol for bipartite qudits called FIMAX demonstrates high performance regarding both distillation efficiency and distillation effectiveness for highly mixed states. The corresponding work \cite{popp_novel_2025} shows that stabilizer operations can be tailored to maximize the increase of fidelity with the maximally entangled state iteratively. Moreover, the action of stabilizer operations on Bell-diagonal states admits a compact description of corresponding output states in terms of the input state, the chosen stabilizer, and the applied encoding. This raises a natural question: can these stabilizer methods be generalized beyond fidelity optimization to enable resource distillation guided by arbitrary information measures, thereby producing states that are highly capable for a broad range of quantum processing tasks?\\
This work develops such a general framework for systems of prime local dimension. We begin in Sec.\ref{sec:setting} by specifying the setting and notation. Sec.\ref{sec:results} develops the main results of this work that enable the stabilizer formalism for general distillation tasks. In particular, in Sec.\ref{sec:stab_rout_channel}, we formalize the stabilizer routine used for entanglement distillation as a quantum channel and extend it to the tripartite setting of cryptographic scenarios, allowing its application to general input states and various distillation tasks. We provide a complete Kraus representation in terms of stabilizer measurements and encodings. Sec.\ref{:sec:principles_stab_resource_dist} demonstrates how these results can be leveraged for general stabilizer-based resource state distillation by optimizing the stabilizer channels for distillation tasks that are characterized by certain information measures. Following the identification of operational simplifications based on invariances of the corresponding information measures in Sec.\ref{sec:simplify}, we
present several protocols to prove the principle of stabilizer-based resource state distillation in Sec.\ref{sec:protocols}. These protocols include a generalization of FIMAX optimizing the fidelity for non-Bell-diagonal states, as well as applications to the (one-shot) entanglement and secret-key distillation by optimizing other information measures. We conclude our results in Sec.\ref{sec:conclusion}. Finally, Sec.\ref{sec:methods} provides a concise summary of preliminary results and established methods used in this work.

\section{Setting and Notation}
\label{sec:setting}
In this section, we briefly introduce the notation relevant for formulating the tasks of stabilizer-based secret-key and entanglement distillation as a quantum channel. For a compact introduction to these methods, the reader is referred to Sec.\ref{sec:methods}, providing more details on stabilizer-based  operations in Sec.\ref{sec:stabBasedDistillation},  the distillation tasks in Sec.\ref{sec:distTasks} and associated information measures in Sec.\ref{sec:subsecBounds}. \\
We consider a setting of two parties Alice ($A$) and Bob ($B$) and the potential adversary third party, Eve ($E$).
\begin{figure}[ht]
    \centering
    \vspace{-1em}
    \includegraphics[width=0.4\linewidth]{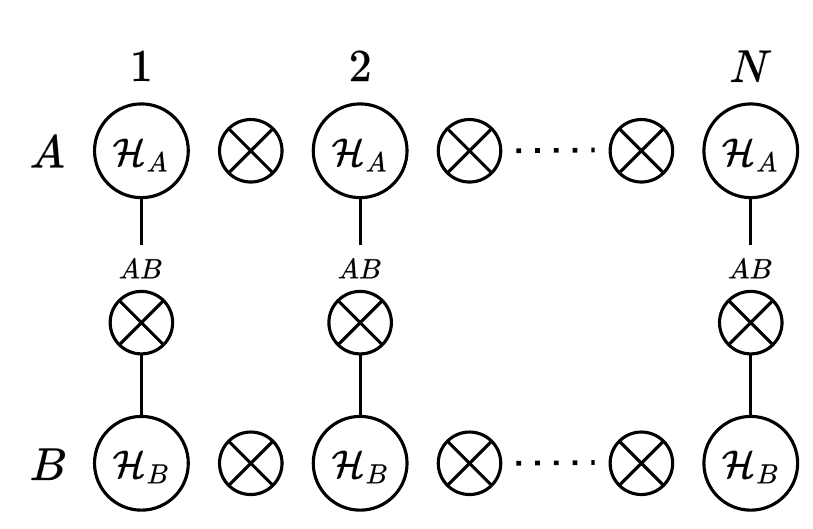}
    \caption{Hilbert space $\hs^{\otimes N}_{AB}$ of $N$-copies of a bipartite system $\hs_{AB} = \hs_A \otimes \hs_B$.}
    \label{fig:hsTensor}
\end{figure}
\noindent Let $\hs_{AB} := \hs_A \otimes \hs_B \cong \mathbbm{C}^d \otimes \mathbbm{C}^d$ be the Hilbert space of bipartite quantum states with dimension of the subsystems $d$. We restrict $d$ to be a prime number. In this case, $\Zd := \mathbbm{Z}/d\mathbbm{Z}$, the quotient ring of integers with addition and multiplication modulo $d$,  is a field. We denote complex conjugation by $(\star)$ and define it in the computational basis for operators.
Let  $\hs^{\otimes N}_{AB} := \bigotimes_{n=1}^N \hs_{AB} \cong \hs_A^{\otimes N} \otimes \hs_B^{\otimes N}$ be the Hilbert space of $N$-copy quantum states as shown in Figure \ref{fig:hsTensor}. 
$\denop(\hs_{AB})$ denotes the set of density operators on $\hs_{AB}$. Pure states and their density matrix are written using the same symbol, e.g., $|\psi\rangle \in \hs \leftrightarrow \psi = |\psi\rangle\langle \psi| \in \denop(\hs)$. Alice and Bob hold $N$ copies of the state $\rho_{AB} \in \denop(\hs_A \otimes \hs_B),~\dim(\hs_A) = \dim(\hs_B) = d$, and aim to establish a target state depending on the distillation task. In the bipartite setting of entanglement distillation, the target is to produce maximally entangled states via LOCC (local operations and classical communication) \eqref{eq:LOCCchannel}, while in the tripartite setting of secret-key distillation, they try to create secret-key states via LOPC (local operations and public classical communication) \eqref{eq:LOPCchannel}. We assume Alice and Bob to hold the state $\rho_{AB}$, while in case the task is secret-key distillation, the third party, Eve, is assumed to possess the purifying system $\hs_E$ for the purification $\phi_{ABE}$ of the state $\rho_{AB}$, giving her a strong position as any extension of $\rho_{AB}$ is related to $\phi_{ABE}$ by a local transformation on her system \cite{khatri_principles_2024}. \\
Any LOCC channel, i.e., completely positive trace-preserving map, used in the bipartite setting of entanglement distillation can be written as
\begin{flalign}
    \label{eq:LOCCchannel}
    \mL_{AB} = \sum_z \mL^z_{A} \otimes \mL^z_{B}  
 \end{flalign}
and acts on the state $\rho_{AB}$. $\mL^z_{A}$ and $\mL^z_{B}$ are completely positive maps, such that the composite map is trace preserving. The corresponding tripartite LOPC channel, used in the setting of secret key-distillation, acts on the purification $\phi_{ABE}$ and reads
\begin{flalign}
    \label{eq:LOPCchannel}
    \mL_{ABE} = \sum_z \mL^z_{A} \otimes \mL^z_{B} \otimes \id_E \otimes |z \rangle\langle z|_E,
\end{flalign} 
considering that any classical communication between Alice and Bob is also known to Eve. 
Quantum states and channels are labeled by the parties that own the systems in the state. If it is not explicitly relevant, we omit specific labels for systems. For example, a state of the systems $A_1$ and $A_2$, both in possession of Alice and formally written $\psi_{A_1, A_2}$, is written as $\psi_{A}$. Marginal states are labeled by a subset of systems, for example $\psi_{AE} := \Tr_B[\psi_{ABE}]$. Since all considered quantum operations are local, we also label the channels with $A,B,E$, according to the parties that perform operations and omit system labels, if possible. For example, a channel mapping the systems $B_1$ and $B_2$ to $E_1$, where $B_1$ and $B_2$ belong to Bob and $E_1$ to Eve, is denoted as $\mL_{BE}$ if the specific systems and the channel direction is not relevant. Otherwise, we write explicitly $\mL_{B_1B_2 \rightarrow E_1}$. Moreover, if a channel, say $\mL_{AB}$ corresponds only to a subset of the parties holding a state, say $\psi_{ABE}$, we identify $\mL_{AB}(\psi_{ABE}) := \mL_{AB}\otimes \textrm{id}_E (\psi_{ABE})$. Isometries are denoted by capital letters $V$, labeled as $V_A$ if only the owner of the systems they act on is relevant, or as $V_{A_1 \rightarrow A_2}$ if the explicit systems the operator acts on are specified. Unitary operators are named by capital $U$. 

\section{Results}
\label{sec:results}
\subsection{The stabilizer routine as a quantum channel}
\label{sec:stab_rout_channel}
In this section, we define the quantum channel corresponding to the stabilizer distillation routine, making it directly available to any bipartite input state and allowing us to extend the distillation setting to a tripartite setting relevant for the distillation of secret-key states.
\subsubsection{The stabilizer distillation routine} 
\label{sec:distRoutine}
In Ref.~\cite{matsumoto_conversion_2003}, a protocol using stabilizer codes for entanglement distillation has been introduced, which has been generalized and analyzed in detail in Ref.~\cite{popp_novel_2025}.
If successful, the recurrently applied bipartite $[N,K]$ distillation protocol transforms $N$ copies of a bipartite input state $\rho_{AB} \in \denop (\hs^{\otimes N}_{AB})$ into $K$ copies of a highly entangled bipartite output state $\sigma^{x,y}_{AB} \in \denop (\hs^{\otimes K}_{AB})$ via LOCC.
For a summary of the formalism for stabilizers and related results, see Sec.\ref{sec:stabBasedDistillation}.\\
Given a stabilizer $S$ with encodings $U(|x\rangle\otimes|k\rangle) = |u_{x,k}\rangle$ and $U^\star(|y\rangle\otimes|l\rangle) = |u^\star_{y,l}\rangle$, $x,y \in \Zd^p$, $k,l \in \Zd^{N-p}$, and corresponding projective stabilizer measurements $ \mP(x,y): \hs^{\otimes N}_{AB} \rightarrow \mQ(x)\otimes \mQ^{\star}(y)$.
One iteration of the protocol includes the following steps:
\begin{enumerate}   
    \item Alice and Bob perform local stabilizer measurements with outcomes $x,y$.
        \item Bob sends Alice his measurement outcome $y$. Alice may declare failure of the protocol depending on $x$ and $y$.
    \item Alice and Bob apply the inverse of stabilizer encodings $U$ (Alice) and $U^\star$ (Bob) on $\hs_A^{\otimes N}$ and $\hs_B^{\otimes N}$.
\end{enumerate}
The state after steps 1-3 can be written as:
\begin{flalign}
    \frac{(U\otimes U^\star)^\dagger~ \mP(x,y) ~\rho_{AB}~ \mP(x,y)~ (U\otimes U^\star)}{\Prob(x,y)} = |x\rangle\langle x| \otimes |y\rangle \langle y| \otimes \sigma^{x,y}_{AB},
\end{flalign}
where the first $p$ qudits contain the measurement outcomes and $\sigma^{x,y}_{AB}$ is the output state to be used in the following iterations. 
Here, $\Prob(x,y) = \Tr[\mP(x,y) ~ \rho_{AB} ]$ corresponds to the probability of obtaining the measurement outcomes $x,y$.

\subsubsection{The bipartite stabilizer channel}
\label{sec:bipartite_setting}
Since, by definition, the codewords $|u_{x,k}\rangle \otimes |u_{y,l}^\star \rangle$  form a basis of the joint codespaces $\mQ(x)\otimes\mQ^{\star}(y)$, we can write 
\begin{flalign}
    \label{eq:stabilizer_measurments}
    \mP_A(x) &= \sum_{k \in \Zd} |u_{x,k} \rangle\langle u_{x,k}| \\
    \mP^\star_B(y) &= \sum_{l \in \Zd} |u^\star_{y,l} \rangle\langle u^\star_{y,l}|\\
    \mP(x,y) &= \sum_{k,l \in \Zd} |u_{x,k}\rangle\langle u_{x,k} | \otimes |u^\star_{y,l}\rangle\langle u^\star_{y,l} |.
\end{flalign}
Now consider the post-measurement state for the input state $\rho_{AB} \in \hs_{AB}^{\otimes N}$ and measurement outcomes $x,y \in \Z_d^{p}$, stored in the additional classical registers $X$ (belonging to Alice) and $Y$ (belonging to Bob), followed by the application of the inverse encoding according to steps 1-3 of the stabilizer distillation routine presented in Sec.\ref{sec:distRoutine}:
\begin{flalign}
    \mL^{x,y}_{AB}(\rho_{AB}) := |x,y\rangle\langle x,y|_{X} \otimes|y\rangle\langle y|_{Y} \otimes \frac{(U\otAB U^\star)^\dagger~ \mP(x,y) ~\rho_{AB}~ \mP(x,y)~ (U\otAB U^\star)}{\Prob(x,y)}.
\end{flalign}
Considering all measurement outcomes, we define the bipartite stabilizer routine channel as:
\begin{flalign}
    \label{eq:bipartite_stabchannel}
    &\mL_{AB}(\rho_{AB}) := \sum_{x,y} |x\rangle\langle x|_{X} \otimes|y\rangle\langle y|_{Y} \otimes(U\otAB U^\star)^\dagger~ \mP(x,y) ~\rho_{AB}~ \mP(x,y)~ (U\otAB U^\star).
\end{flalign}
Introducing the operators
\begin{flalign}
    \label{eq:kraus_ops}
    M^x_A &:= \sum_k (|x\rangle \otimes |k\rangle)\langle u_{x,k}| = U^\dagger \mP_A(x), \hspace{13em} (M^x_A: \hs_A^{\otimes N} \rightarrow \hs_A^{\otimes N}), \\
    \label{eq:kraus_ops_2}
    M^{\star y}_B &:= \sum_l (|y\rangle \otimes |l\rangle)\langle u^\star_{y,l}| = (U^{\star})^{\dagger} \mP^\star_B(y),\hspace{12.2em}(M^{\star y}_B:\hs_B^{\otimes N} \rightarrow \hs_B^{\otimes N}), \\
    K_{AB}^{x,y} &:= |x\rangle_X \otimes |y\rangle_Y \otimes M_A^x \otimes M_B^{\star y}, \hspace{14.5em}(K_{AB}^{x,y}:\hs_{AB}^{\otimes N} \rightarrow \hs_{XY} \otimes \hs_{AB}^{\otimes N}),
\end{flalign}
we have:
\begin{flalign}
    \mathcal{L}_{AB}(\rho_{AB}) &= \sum_{x,y} |x\rangle\langle x|_{X} \otimes|y\rangle\langle y|_{Y} \otimes (M_A^x \otimes M_B^{\star y})~\rho_{AB}~(M_A^{x \dagger} \otimes M_B^{y \dagger})
    = \sum_{x,y} K_{AB}^{x,y} ~\rho_{AB}~ K_{AB}^{x,y\dagger}.
\end{flalign}
Note that $\sum_{x,y} K_{AB}^{x,y\dagger} K_{AB}^{x,y} = \mathbbm{1}_{AB}$. Hence, the $K_{AB}^{x,y}$ are Kraus operators for the quantum channel $\mL_{AB}$. We call this channel the ``bipartite stabilizer channel''. \\
For fixed measurement outcomes $x,y$, the output state $\sigma^{x,y}_{AB}:=\mathcal{L}^{x,y}_{AB}(\rho_{AB})$ can be written in terms of $K_{AB}^{x,y}$ as follows:
\begin{flalign}
    \label{eq:output_state_kraus}
    \sigma_{AB}^{x,y} = \frac{K_{AB}^{x,y} ~\rho_{AB}~ K_{AB}^{x,y \dagger}}{\Tr[K_{AB}^{x,y} ~\rho_{AB}~ K_{AB}^{x,y \dagger}]} = \frac{1}{\Prob(x,y)}K_{AB}^{x,y} ~\rho_{AB}~ K_{AB}^{x,y \dagger}.
\end{flalign}

\subsubsection{The tripartite stabilizer channel}
\label{se:tripartite}
In the setting of quantum cryptography, e.g., for the distillation or distribution of secret keys, we consider a tripartite setting by including the adversary party ``Eve'' holding the system $E$. We assume that Eve holds a purification $\phi_{ABE}$ of the state $\rho_{AB}$ and that she can perform any local operation on her system. Additionally, public classical communication is assumed. So any classical information sent between Alice and Bob is also known to Eve, and an LOCC channel becomes an LOPC (local operations and public classical communication) channel. In the following, we assume that all measurement outcomes are communicated in public. Consider the purification based on the spectral decomposition of $\rho_{AB}$ in terms of its eigenstates $\lbrace |\Phi_m \rangle \rbrace_{m \in \lbrace 0, \dots, r-1 \rbrace}$ with $r = \mathrm{rank}(\rho_{AB})$ and nonzero eigenvalues $\lambda_m$:
\begin{flalign}
    \label{eq:purification}
    \rho_{AB} &= \sum_m \lambda_m | \Phi_m \rangle\langle \Phi_m |_{AB} \\
    \phi_{ABE} &= |\phi\rangle\langle\phi|_{ABE} = \sum_{m,n} \sqrt{\lambda_m \lambda_n} |\Phi_m \rangle \langle \Phi_n|_{AB} \otimes | m \rangle\langle n |_E.
\end{flalign}
Here $\lbrace |m\rangle \rbrace_m$ is a basis of the purifying system with dimension $r$.
The tripartite channel corresponding to the bipartite stabilizer distillation channel is defined as:
\begin{flalign}
    \label{eq:tripartite_stabchannel}
    \mathcal{L}_{ABE}(\phi_{ABE}) &:= \mathcal{L}_{AB} \otimes  \textrm{id}_E (\phi_{ABE}) \\
    &= \sum_{x,y} K_{AB}^{x,y} \otimes \mathbbm{1}_E ~\phi_{ABE}~ K_{AB}^{x,y \dagger} \otimes \mathbbm{1}_E,
\end{flalign}
from which we read off the Kraus operators $\hat{K}_{x,y} = K_{AB}^{x,y} \otimes \mathbbm{1}_E$. We call this channel the ``tripartite stabilizer channel''.\\
Let $ \lbrace |\Omega(e)\rangle \rbrace_{e \in \Zd^N}$ be the $N$-copy Bell basis (cf.~Def.\ref{def:bell_states}) of $\hs_{AB}^{\otimes N}$. 
Writing $|\Phi_m\rangle\langle \Phi_n |_{AB} = \sum_{e,f} \beta_{e,f}^{m,n} ~|\Omega(e) \rangle \langle \Omega(f)|_{AB}$ in this basis, $\phi_{ABE}$ takes the form:
\begin{flalign}
    \label{eq:tripartite_output_alpha_form}
    \phi_{ABE} &= \sum_{m,n=0}^{r-1} \sum_{e,f \in \Zd^N} \sqrt{\lambda_m \lambda_n}~ \beta_{e,f}^{m,n} ~|\Omega(e) \rangle \langle \Omega(f)|_{AB} \otimes |m \rangle \langle n|_E \\
    &=: \sum_{\substack{m,n,\\e,f}}  \alpha_{e,f}^{m,n} ~|\Omega(e) \rangle \langle \Omega(f)|_{AB} \otimes |m \rangle \langle n|_E.
\end{flalign}
For a Bell-diagonal input state $\rho_{AB} = 
\sum_m  ~p_m | \Omega(m) \rangle \langle \Omega(m) |$ this simplifies to 
\begin{flalign}
    \phi_{ABE} = \sum_{m,n = 0}^{r-1} \sqrt{p_m p_n} ~| \Omega(m) \rangle \langle \Omega(n) |_{AB} \otimes | m \rangle \langle n |_E. 
\end{flalign}
In Ref.\cite{popp_novel_2025} it is demonstrated that the action of stabilizer operations on Bell states $|\Omega(e)\rangle$ can be specified by so-called action operators $T^{U,e}_x$, of dimension $d^{\otimes N-p}$ and characterized by (cf. Lemma \ref{thm:encoding_errors})
\begin{flalign}
\label{eq:action_ops}
U^\dagger W(e) U = \sum_{x\in \Zd^p} |x+s \rangle \langle x | \otimes T_{x+s}^{U,e}.
\end{flalign} 
Here, $s=s(e) \in \Z_d$ is the ``syndrome'' of $e$ and depends on the stabilizer and the error $e$. The unitary operators $T$ depend on the encoding $U$, the error $e$, and the measurement outcome $x$. Assuming the encoding $U$ to be fixed, we omit the label and write simply $T^e_x$. \\
The form of the operators $M^x_A = \sum_k (|x\rangle \otimes|k\rangle)\langle u_{x,k}|$ and $M^{*y}_B=\sum_l (|y\rangle\otimes|l\rangle)\langle u^{\star}_{y,l}|$ \eqref{eq:kraus_ops}-\eqref{eq:kraus_ops_2} together with \eqref{eq:action_ops} now imply:
\begin{flalign}
    M^x_A\otimes M^{*y}_B ~|\Omega(e)\rangle &= (M^x_A \otimes M^{*y}_B)~(W(e)\otimes \mathbbm{1})~(U\otimes U^\star) ~|\Omega_{0,0}\rangle^{\otimes N} \\
    &= (|x\rangle \langle x-s| \otimes T_x^e) \otimes (|y \rangle \langle y|) \otimes \id) ~|\Omega_{0,0}\rangle^{\otimes N}\\
    &= \frac{1}{d^{p/2}} \delta_{x-y,s}~(|x\rangle \otimes |y\rangle)  \otimes (T_x^{e} \otimes \mathbbm{1}) ~ |\Omega_{0,0}\rangle^{\otimes N-p}.
\end{flalign}
One then finds for the output of the tripartite stabilizer distillation channel for the input of \eqref{eq:tripartite_output_alpha_form}:
\begin{flalign}
    \label{eq:tripartite_stabchannel_explicit}
    \mathcal{L}_{ABE}(\phi_{ABE}) 
    &= \sum_{\substack{x,y,\\m,n,\\e,f \in \E(x-y)}} 
    \alpha_{e,f}^{m,n}~\frac{1}{d^p}
    |x,y \rangle \langle x,y |_{XYZ} \otimes [(T^e_x \otimes \mathbbm{1}) ~\Omega_{0,0}^{\otimes N-p}~ (T_x^{f\dagger} \otimes \mathbbm{1})]_{AB} \otimes | m \rangle \langle n|_E.
\end{flalign}
For fixed measurement outcomes $x,y$, stored in the classical registers $X,Y,Z$ of Alice, Bob, and Eve, the tripartite post-measurement output state of only the non-classical systems $A,B,E$ is:
\begin{flalign}
    \label{eq:tripartite_output_state}
    \tau^{x,y}_{ABE} &= \langle x,y|_{XYZ} \otimes \id_{ABE}~
    \left( \frac{\hat{K}_{x,y} ~ \phi_{ABE} ~\hat{K}_{x,y}^\dagger}{\Tr[\hat{K}_{x,y} ~ \phi_{ABE} ~\hat{K}_{x,y}^\dagger]} \right)
    ~ |x,y\rangle_{XYZ} \otimes \id_{ABE} \\
    &= \frac{1}{\Prob(x,y)}~\sum_{\substack{m,n,\\e,f \in \E(x-y)}}  
    \alpha_{e,f}^{m,n}\frac{1}{d^p}~[(T^e_x \otimes \mathbbm{1}) ~\Omega_{0,0}^{\otimes N-p}~ (T_x^{f\dagger} \otimes \mathbbm{1})]_{AB} \otimes | m \rangle \langle n|_E,
\end{flalign}
where we used the fact that the only trace-decreasing operation in $\mathcal{L}_{ABE}$ for fixed $x,y$ is the bipartite stabilizer measurement via $\mP(x,y)$ and that $\phi_{ABE}$ is the purification of $\rho_{AB}$, implying $
    \Prob(x,y) = \Tr[\hat{K}_{x,y} ~ \phi_{ABE} ~\hat{K}_{x,y}^\dagger].$
\subsection{Principle and relevant objects for stabilizer-based resource distillation}
\label{:sec:principles_stab_resource_dist}
In this section we show how to optimize the stabilizer channels so that their bipartite and tripartite output states are highly potent for operational tasks like entanglement and secret-key distillation. \\
The capability of general quantum states for the entanglement and secret-key distillation tasks can be characterized or lower-bounded via generalized information quantities. We summarize methods of how to do so in Sec.\ref{sec:distTasks} and Sec.\ref{sec:subsecBounds}.
The principle of stabilizer-based resource state distillation is as follows: depending on the distillation task, we choose an information quantity that measures the corresponding capability of a quantum state. Leveraging the precise output form of the stabilizer channels, we optimize the channel regarding the stabilizer, encoding, and measurement outcome such that the information measure for the channel output becomes maximized. This procedure can be iterated, effectively distilling target resource states that are optimal for the operational task as measured by the chosen information quantity. In the following, we determine relevant states and information measures for entanglement and secret-key distillation and leverage their properties to simplify the optimization of the distillation channel.

\subsubsection{Relevant states and measures for entanglement distillation}
\label{sec:ent_dist_relevant_states}
Tracing out the system $E$ from the tripartite output state \eqref{eq:tripartite_output_state}, the general bipartite output state \eqref{eq:output_state_kraus} takes the specific form:
\begin{flalign} 
    \label{eq:bipartite_output_alpha_form}
        \sigma_{AB}^{x,y} &= \frac{1}{\Prob(x,y)}K_{AB}^{x,y} ~\rho_{AB}~ K_{AB}^{x,y \dagger} \notag \\
        &= \frac{1}{\Prob(x,y)}~\sum_{\substack{m \\e,f \in \E(x-y)}}  
        \alpha_{e,f}^{m,m}\frac{1}{d^p}~[(T^e_x \otimes \mathbbm{1}) ~\Omega_{0,0}^{\otimes N-p}~ (T_x^{f\dagger} \otimes \mathbbm{1})]_{AB}.
\end{flalign}
In the one-shot setting, we leverage Theorem \ref{thm:lowerboundOSEnt} to determine the smooth max-relative entropy $H^{\varepsilon}_{max}(\sigma_{AB}^{x,y})$ as an information measure to be optimized for distilling resource states with a high capability for entanglement distillation. \\
In the asymptotic setting, Theorem \ref{thm:lowerboundentAS} implies that the coherent information $I^{A\rangle B}_c(\sigma^{x,y}_{AB})$ is an achievable rate for entanglement distillation and therefore provides a suitable information measure for optimization.

\subsubsection{Relevant states and measures for secret-key distillation}
\label{sec:key_dist_relevant_states}
For the task of secret-key distillation, consider the output \eqref{eq:tripartite_output_state} of the tripartite stabilizer distillation channel, given the outcome $x,y$:
\begin{flalign}
    \tau^{x,y}_{ABE} : = \frac{1}{\Prob(x,y)} 
    \sum_{\substack{m,n, \\ e,f \in \E(x-y)}}
    \alpha_{e,f}^{m,n} [(T^e_x \otimes \mathbbm{1}_B) ~\Omega_{0,0}^{\otimes N-p}~ (T_x^{f\dagger} \otimes \mathbbm{1}_B)]_{AB} \otimes | m \rangle \langle n |_E.
\end{flalign}
To draw a key from this state, Alice and Bob apply local measurements to their system, whose action is given by the local measurement or dephasing channels $\mathcal{D}_{A/B}(\cdot)$ in \eqref{eq:measurement_channels}.
Assume that Alice applies $\mathcal{D}_A$ to the output state:
\begin{flalign}
    \label{eq:tripartie_locdephased_state} 
    \chi^{x,y}_{ABE} := \mathcal{D}_A ( \tau^{x,y}_{ABE}) 
    = \frac{1}{\Prob(x,y)} 
    \sum_{\substack{m,n, \\ e,f \in \E(x-y), \\ i,k,l \in \Zd^{N-p}}}
    \alpha^{m,n}_{e,f}~\langle i | T_x^e | k \rangle \langle l | T_x^{f\dagger} | i \rangle 
    ~|i\rangle \langle i|_A \otimes | k \rangle \langle l |_B \otimes | m \rangle \langle n |_E. 
\end{flalign}
To evaluate this state regarding its usefulness for secret-key generation, we consider the reduced bipartite states:
\begin{flalign}
    \label{eq:bibartite_locdephased_AB}
    \chi^{x,y}_{AB} &:= \Tr_E[\chi_{ABE}] = \frac{1}{\Prob(x,y)} 
    \sum_{\substack{m, \\ e,f \in \E(x-y), \\ i,k,l \in \Zd}}
    \alpha^{m,m}_{e,f}~\langle i | T_x^e | k \rangle \langle l | T_x^{f\dagger} | i \rangle 
    ~|i\rangle \langle i|_A \otimes | k \rangle \langle l |_B,
\end{flalign}
\begin{flalign}    
    \label{eq:bibartite_locdephased_AE}
    \chi^{x,y}_{AE} &:= \Tr_B [\chi_{ABE}] = \frac{1}{\Prob(x,y)} 
    \sum_{\substack{m,n, \\ e,f \in \E(x-y), \\ i \in \Zd}}
    \alpha^{m,n}_{e,f}~\langle i | T_x^e  T_x^{f\dagger} | i \rangle ~|i\rangle \langle i|_A \otimes | m \rangle \langle n|_E.
\end{flalign}
In the one-shot setting, Theorem \ref{thm:lowerboundOSS} implies that the difference of the hypothesis testing mutual information $I^\varepsilon_H(\chi^{x,y}_{AB})$ and the smooth max-mutual information $I^\epsilon_{max}(\chi^{x,y}_{AE})$ provides achievable rates for the one-shot secret-key distillation, given certain smoothing parameters $\varepsilon$ and $\epsilon$. Optimization of this difference by varying $\varepsilon$, $\epsilon$, and the channel output $\tau^{x,y}_{ABE}$ (determining $\chi_{AB}$ and $\chi_{AE}$) hence optimizes the resource for one-shot secret key distillation.\\
Likewise, in the asymptotic setting, the so-called private information $I(\chi_{ABE}):= I(\chi_{AB}) - I(\chi_{AE})$ provides an achievable rate according to Theorem \ref{thm:lowerboundkeyAS} and serves as a quantity to be optimized via stabilizer channels.

\subsection{Simplifications for local isometric and unitary invariance}
\label{sec:simplify}
In Ref.\cite{popp_local_2025}, it is shown that all considered divergence-based measures, as discussed above and defined in Sec.\ref{sec:methods}, are invariant under local isometric or unitary transformations.
We leverage this property in two ways. First, we show that a different choice of encoding for the stabilizer channel corresponds to a local unitary transformation of the output state. Hence, given a certain information measure, the choice of encoding is irrelevant for finding the optimal stabilizer channel if the information measure is invariant. Second, we show that the tripartite purification of the bipartite stabilizer channel output is related to the output of the tripartite channel by a local isometric transformation. Given a local isometric invariant information measure in the tripartite distillation setting, the purification of the bipartite channel output can be considered instead of the tripartite channel output. This effectively reduces the dimension of the states to evaluate and therefore brings significant performance improvements for calculating the information measure for distillation.

\subsubsection{Local unitary equivalence of encodings}
\label{sec:simplify_encoding}
We now show that a change of stabilizer encoding implies a unitary transformation of the bipartite output state, acting locally on the systems $A$ and $B$. Following Proposition 6 of Ref.~\cite{popp_novel_2025}, given a stabilizer encoding $U$, for any other stabilizer encoding $V$, there exist unitary matrices $Z_k$, such that 
$V = U~(\sum_k |k \rangle\langle k| \otimes Z_k)$ and $T^{V,e}_x = Z_x^\dagger T^{U,e}_x Z_y$. This implies for the output state \eqref{eq:bipartite_output_alpha_form} using the encoding $V$.
\begin{flalign}
    \label{eq:bilocal_unitary_encoding_action}
    \sigma^{x,y,V}_{AB} &= \frac{1}{\Prob(x,y)} \sum_{\substack{m \\e,f \in \E(x-y)}}   \alpha_{e,f}^{m,m}\frac{1}{d^p}~ (Z_x^\dagger T^{U,e}_x Z_y \otimes \mathbbm{1}) 
    ~\Omega^{\otimes N-p}_{0,0}~ 
    (Z_x^\dagger T^{U,f}_x Z_y \otimes \mathbbm{1})^\dagger \notag  \\
    &= \frac{1}{\Prob(x,y)} \sum_{\substack{m \\e,f \in \E(x-y)}}   \alpha_{e,f}^{m,m}\frac{1}{d^p}~ 
    (Z_x^\dagger T^{U,e}_x \otimes Z_y^T) 
    ~\Omega^{\otimes N-p}_{0,0}~ 
    (Z_x^\dagger T^{U,f}_x \otimes Z_y^T)^\dagger \notag \\
    &= \frac{1}{\Prob(x,y)} \sum_{\substack{m \\e,f \in \E(x-y)}}   \alpha_{e,f}^{m,m}\frac{1}{d^p}~ 
    (Z_x^\dagger \otimes Z_y^T)
    (T^{U,e}_x \otimes \mathbbm{1}) 
    ~\Omega^{\otimes N-p}_{0,0}~ 
    (T^{U,f}_x \otimes \mathbbm{1})^\dagger
    (Z_x \otimes Z_y^\star) \notag \\
    &= (Z_x^\dagger \otimes Z_y^T) ~\sigma^{x,y,U}_{AB}~ (Z_x \otimes Z_y^\star).
\end{flalign}
Note that the coefficients $\alpha$ only depend on the input state and that the probabilities $\mP(x,y)$ only depend on the codespaces $\mQ(x) \otimes \mQ(y)$ and not on the choice of encoding, i.e., their basis. 
Also, for the second equation, we used the ``transpose ($T$) trick'', holding for any matrix $M \in \mathbb{C}^{d_A \times d_B}$: $M \otimes \id |\Omega_{0,0}\rangle = \id \otimes M^T |\Omega_{0,0}\rangle$.\\
Equivalently, one finds for the tripartite channel output:
\begin{flalign}
    \label{eq:tripartite_unitary_encoding_action}
    \tau^{x,y,V}_{ABE} &= (Z_x^\dagger \otimes Z_y^T)_{AB} ~\tau^{x,y,U}_{ABE}~ (Z_x \otimes Z_y^\star)_{AB}.
\end{flalign}
Consequently, any information measure that is invariant under local unitary transformations is invariant under change of encodings for the stabilizer channel when measuring the channel output.

\subsubsection{Local isometric equivalence of purifications}
\label{sec:simplify_purification}
We now show that the tripartite purification $\psi^{x,y}_{ABE'}$ of the bipartite output state $\sigma^{x,y}_{AB}$ of the bipartite channel $\mathcal{L}_{AB}$ for a certain measurement outcome $x,y$ is related to the post-measurement output state $\tau^{x,y}_{ABE}$ of the tripartite channel $\mathcal{L}_{ABE}$ applied to the purification $\phi_{ABE}$ of the input state $\rho_{AB}$. \\
By \eqref{eq:bipartite_stabchannel} and \eqref{eq:tripartite_stabchannel}, the channel $\mathcal{L}_{ABE}$ consists of a projective measurement followed by a unitary transformation. Given the purification of the input state $\phi_{ABE}$, applying $\mathcal{L}_{ABE}$ and projecting to the measurement outcomes $x,y$, the resulting state $\tau^{x,y}_{ABE}$ is also a pure state. 
We have for the purification $\psi^{x,y}_{ABE'}$ of the bipartite output state $\sigma^{x,y}_{AB}$ and the tripartite output state $\tau^{x,y}_{ABE}$:
\begin{flalign}
    \Tr_{E'}[\psi^{x,y}_{ABE'}] = \sigma^{x,y}_{AB} = \Tr_E[\tau^{x,y}_{ABE}]. 
\end{flalign} 
Hence, both $\psi^{x,y}_{ABE'}$ and $\tau^{x,y}_{ABE}$ are purifications of $\sigma^{x,y}_{AB}$ and as such related by an isometry $V_{E' \rightarrow E}$ \cite{khatri_principles_2024}, mapping one purification system to the other: 
\begin{flalign}
    \tau^{x,y}_{ABE} = V_{E' \rightarrow E}~\psi^{x,y}_{ABE'}~ V^\dagger_{E' \rightarrow E}.
\end{flalign}
As the isometry only acts on the purifying system, also the locally dephased states satisfy this relation:
\begin{flalign}
    \label{eq:equivalent_dephased_state}
    \chi^{x,y}_{ABE} &= \mathcal{D}_A(\tau^{x,y}_{ABE}) \notag \\
    &= \mathcal{D}_A(V_{E' \rightarrow E} ~\psi^{x,y}_{ABE'}~ V^\dagger_{E' \rightarrow E}) \notag \\
    &= V_{E' \rightarrow E} ~\mathcal{D}_A (\psi^{x,y}_{ABE'})~ V^\dagger_{E' \rightarrow E} \notag \\
    &=: V_{E' \rightarrow E} ~\zeta^{x,y}_{ABE'}~ V^\dagger_{E' \rightarrow E} ,
\end{flalign}
where we defined the partially dephased state $\zeta^{x,y}_{ABE'} = \mathcal{D}_A (\psi^{x,y}_{ABE'})$ based on the purification $\psi^{x,y}_{ABE'}$.
Consequently, the reduced states $\chi_{AB}$ \eqref{eq:bibartite_locdephased_AB} and $\chi_{AE}$ \eqref{eq:bibartite_locdephased_AE} satisfy
\begin{flalign}
    \label{eq:equivalent_dephased_reduced_states}
    \chi^{x,y}_{AB} &= \zeta^{x,y}_{AB}  \\
    \chi^{x,y}_{AE} &= V_{E' \rightarrow E} ~\zeta^{x,y}_{AE'}~ V^\dagger_{E' \rightarrow E} \label{eq:equivalent_dephased_reduced_states_AE}.
\end{flalign}
Note that because $E'$ is a purification system for the $d^{N-p}$-dimensional output state $\sigma^{x,y}_{AB}$, $E'$ can be chosen to be $d^{2(N-p)}$-dimensional if $\sigma^{x,y}_{AB}$ has full rank. The system $E$ used for the purification of the $d^N$-dimensional input state $\rho_{AB}$, however, needs to be $d^{2N}$-dimensional if $\rho_{AB}$ has full rank. Hence, considering states based on $E'$ has computational advantages, for the calculation of quantities that are invariant under isometric transformations on the purifying system. Since the Hilbert space dimension grows exponentially with the number of qudits, this reduction lowers the computational complexity of stabilizer-based protocols—such as the SPI-IMAX protocol introduced below—by orders of magnitude, transforming otherwise intractable optimizations into feasible numerical tasks. 

\subsection{Defining stabilizer-based resource distillation protocols}
\label{sec:protocols}
In this section, we define specific protocols that demonstrate the principle of stabilizer-based resource distillation. We consider the operational tasks of entanglement distillation (cf. Sec.\ref{sec:ed_task}) and secret-key distillation (cf. Sec.\ref{sec:skd_task}) and define protocols that optimize corresponding information measures via stabilizer operations.

\subsubsection{Entanglement distillation}
\label{sec:specific_ent_protocols}
\begin{figure}[H]
    \vspace{-1em}
    \centering
    \includegraphics[width=0.8\linewidth]{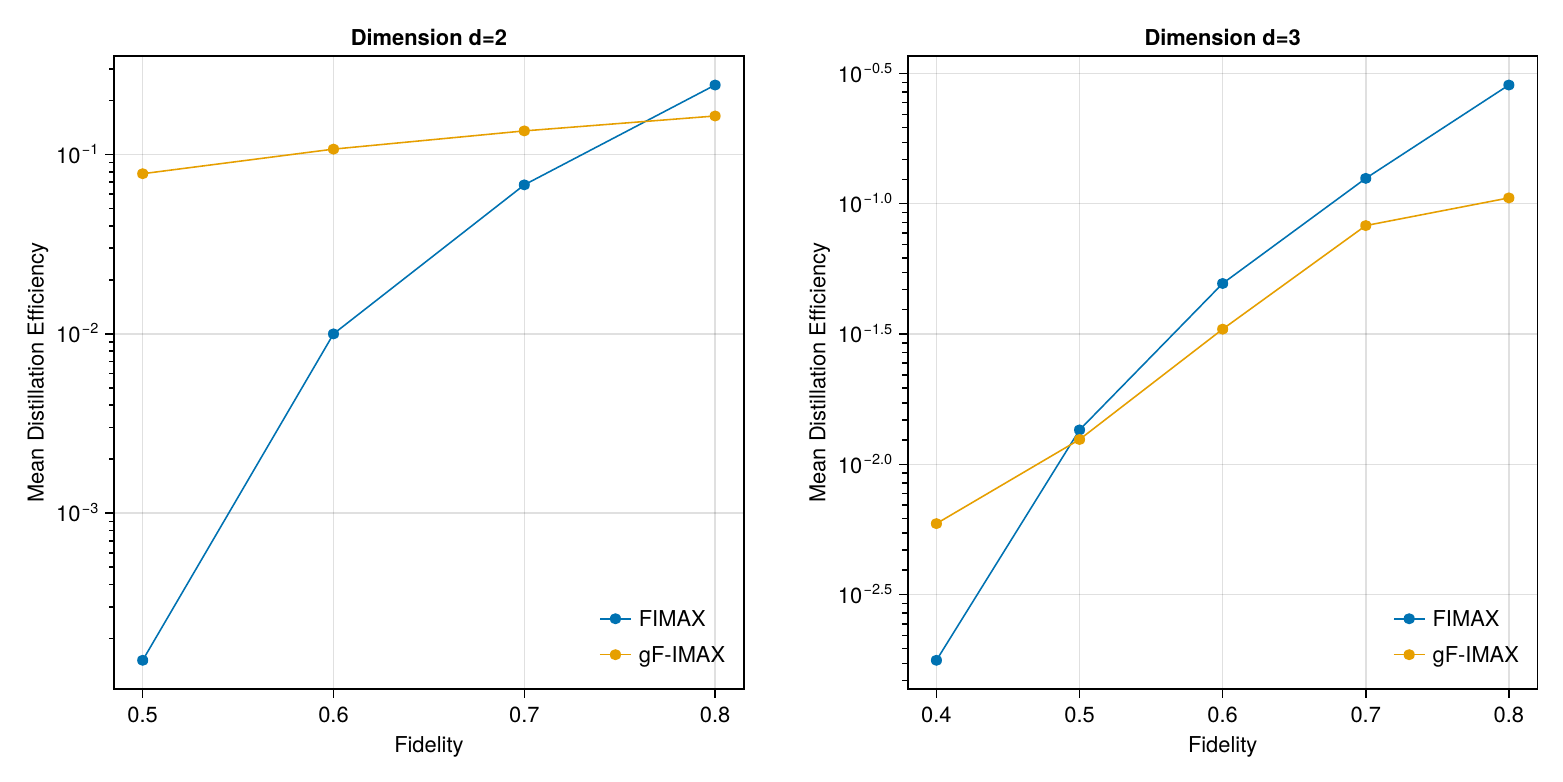} 
    \caption{
    Protocol comparison of mean distillation efficiencies to reach a target fidelity of $0.9$ depending on the fidelity of pure states. Each data point represents the mean efficiency for a group of $1000$ bipartite entangled input states with fidelities rounded to one digit for local dimensions $d=2$ (left) and $d=3$ (right). The efficiency is defined as the inverse of the required number of input states to produce one output state with fidelity larger than the target fidelity. To generate these results, we used the methods implemented in Ref.\cite{popp_belldiagonalqudits_2023}.
    }
    \label{fig:pur_eff_gFimax}
\end{figure}

The first example extends the recently introduced FIMAX protocol \cite{popp_novel_2025}. Using a specific ``canonical encoding $U_c$'', \mbox{FIMAX} requires Bell-diagonal states to increase the fidelity with the maximally entangled state $\mathcal{F}(\rho):= \langle \Omega_{0,0} | ~\rho ~| \Omega_{0,0}\rangle$. For two copies of the input state and prime dimension, FIMAX is proven to maximize the fidelity increase per iteration for Bell-diagonal states among stabilizer-based distillation protocols. Leveraging the general output of the bipartite stabilizer channel $\sigma^{x,y}_{AB}$, we define the ``generalized Fidelity Increase Maximizing (gF-IMAX)'' protocol that is applicable for all two-copy input states in prime dimension. \\ \ \\
\textbf{gF-IMAX protocol:}\\
Let $\rho_{AB}$ be any bipartite state shared between Alice and Bob with prime dimensions $\dim(\hs_A) = \dim(\hs_B) = d$ and let $U_c$ be the canonical encoding (cf. Ref.\cite{popp_novel_2025}) for two copies of $\rho_{AB}$. The following steps are recurrently applied.
\begin{enumerate}
    \item For each stabilizer $S$, all measurement outcomes $x,y \in \Zd$ and all $k,l \in \Zd$, Alice computes $\mathcal{F}_{k,l}(\sigma^{x,y}_{AB}) = \langle \Omega_{k,l} | ~ \sigma^{x,y}_{AB}~| \Omega_{k,l}\rangle $ for the corresponding post-measurement stabilizer channel outputs $\sigma^{x,y}_{AB}$ as in \eqref{eq:bipartite_output_alpha_form}.
    \item Alice determines the configuration $(S_{max}, x_{max}, y_{max}, k_{max}, l_{max})$ that maximizes $\mathcal{F}_{k,l}(\sigma^{x,y}_{AB}))$ and communicates it to Bob.
    \item Alice and Bob apply the stabilizer channel corresponding to $S_{max}$ with the canonical encoding $U_c$ and determine their outcomes $x,y$. If $x \neq x_{max}$ or $y \neq y_{max}$, they declare failure of the protocol. 
    \item Alice applies the unitary Weyl-Heisenberg transformation $W_{k_{max},l_{max}}^\dagger ~(\cdot)~W_{k_{max},l_{max}}$ (cf. Def.\ref{def:weylOpsErrs}) locally to her output.
\end{enumerate}
Although both increase the fidelity by using the stabilizer operations with the canonical encoding, there are relevant differences between gF-IMAX and FIMAX. First, unlike FIMAX for Bell-diagonal states, the canonical encoding does not imply a maximal increase of fidelity in each iteration among all stabilizers, encodings, and measurement outcomes. Depending on the input state, some other encoding may achieve better performance. Second, the condition for failure of FIMAX is given by the difference of measurement outcomes $s:=x-y \neq s_{max}$. This is because for Bell-diagonal states, the channel output only depends on the syndrome, i.e., the difference of measurement outcomes $s = x-y$, which does not hold for general input states. gF-IMAX fails unless precisely $x = x_{max}$ an $y = y_{max}$, implying a higher probability of failure due to stricter post-selection, and thus a negative impact on the protocol efficiency. On the other hand, to apply FIMAX to non Bell-diagonal states, a twirl to Bell-diagonal form is required, generally inducing additional noise. As this is not the case for gF-IMAX, it may be more efficient for non Bell-diagonal states. This is demonstrated in Figure \ref{fig:pur_eff_gFimax}, where the efficiency of distilling states is shown in dependence on the input fidelity of random pure states. The performance of gF-IMAX exceeds FIMAX for certain low-fidelity regions, while FIMAX is superior for higher fidelities.
\noindent For more general resource distillation protocols, we consider other quantities than the fidelity to optimize with stabilizer channels.
Theorem \ref{thm:lowerboundOSEnt} states that the negative smooth conditional max-entropy $(-H^\varepsilon_{max})$ (Def.\ref{def:maxrelEntInf}), also denoted as ``smooth coherent information'', defines a lower bound on the one-shot distillable entanglement. We want to optimize this quantity via stabilizer operations. According to Ref.\cite{popp_local_2025}, $H^\varepsilon_{max}(\rho_{AB})$ is invariant under local unitary transformation of the bipartite state $\rho_{AB}$. Due to the results of Sec.\ref{sec:simplify_encoding}, all channels for some stabilizer with different encodings $U$ produce output states $\sigma^{x,y,U}_{AB}$ with equal smooth conditional max-entropy $H^\varepsilon_{max}(\sigma^{x,y,U}_{AB})$. To maximize the smooth coherent information $(-H^{\varepsilon}_{max})$, it consequently suffices to consider different stabilizers with fixed arbitrary encodings.
We define the ``Smooth Coherent Information Increase Maximizing (SCI-IMAX)'' protocol as follows.\\ \ \\
\textbf{SCI-IMAX protocol}:\\
Let $\rho_{AB}$ be any bipartite state shared between Alice and Bob with prime local dimensions $d_A = d_B$ and $\varepsilon \in (0,1)$. The following steps are recurrently applied.
\begin{enumerate}
    \item For each stabilizer $S$ Alice and Bob fix an arbitrary encoding $U_S$.
    \item For all $S$ and all measurement outcomes $x,y$, Alice computes $H^{\varepsilon}_{max}(\sigma^{x,y}_{AB})$ for the corresponding post-measurement stabilizer channel outputs $\sigma^{x,y}_{AB}$ \eqref{eq:bipartite_output_alpha_form}.
    \item Alice determines the configuration $(S_{max}, x_{max}, y_{max})$ that maximizes $(-H^{\varepsilon}_{max}(\sigma^{x,y}_{AB}))$ and communicates it to Bob.
    \item Alice and Bob apply the stabilizer channel corresponding to $S_{max}$ with the arbitrary encoding $U_{S_{max}}$ and determine their outcomes $x,y$. If $x \neq x_{max}$ or $y \neq y_{max}$, they declare failure of the protocol. 
\end{enumerate}
According to Theorem  \ref{thm:lowerboundentAS}, the coherent information $I^{A \rangle B}_c$ (Def.\ref{def:quantRelEnt}) provides a lower bound for the distillable entanglement in the asymptotic setting. This quantity is also invariant under local unitaries \cite{popp_local_2025} and consequently under a change of encodings. We therefore define the ``Coherent Information Increase Maximizing (CI-IMAX)'' protocol. \\ \ \\ \ \\
\textbf{CI-IMAX Protocol:}\\
Let $\rho_{AB}$ be any bipartite state shared between Alice and Bob with prime local dimensions $d_A = d_B$. The following steps are recurrently applied.
\begin{enumerate}
    \item For each stabilizer $S$ Alice and Bob fix an arbitrary encoding $U_S$.
    \item For all $S$ and all measurement outcomes $x,y$, Alice calculates $I^{A\rangle B}_{c}(\sigma^{x,y}_{AB})$ for the corresponding post-measurement channel outputs $\sigma^{x,y}_{AB}$ \eqref{eq:bipartite_output_alpha_form}.
    \item Alice determines the configuration $(S_{max}, x_{max}, y_{max})$ that maximizes $I^{A \rangle B}_c(\sigma^{x,y}_{AB})$ and communicates it to Bob.
    \item Alice and Bob apply the stabilizer channel corresponding to $S_{max}$ with the arbitrary encoding $U_{S_{max}}$ and determine their outcomes $x,y$. If $x \neq x_{max}$ or $y \neq y_{max}$, they declare failure of the protocol. 
\end{enumerate}
Figure \ref{fig:resource_dist_example} shows how the coherent information and the smooth coherent information are increased by applying the corresponding protocols to an isotropic input state. For the numericals, the implementation of \emph{BellDiagonalQudits.jl} \cite{popp_belldiagonalqudits_2023} is used.

\subsubsection{Secret-key distillation}
\begin{figure}[H]
    \centering
    \includegraphics[width=1\linewidth]{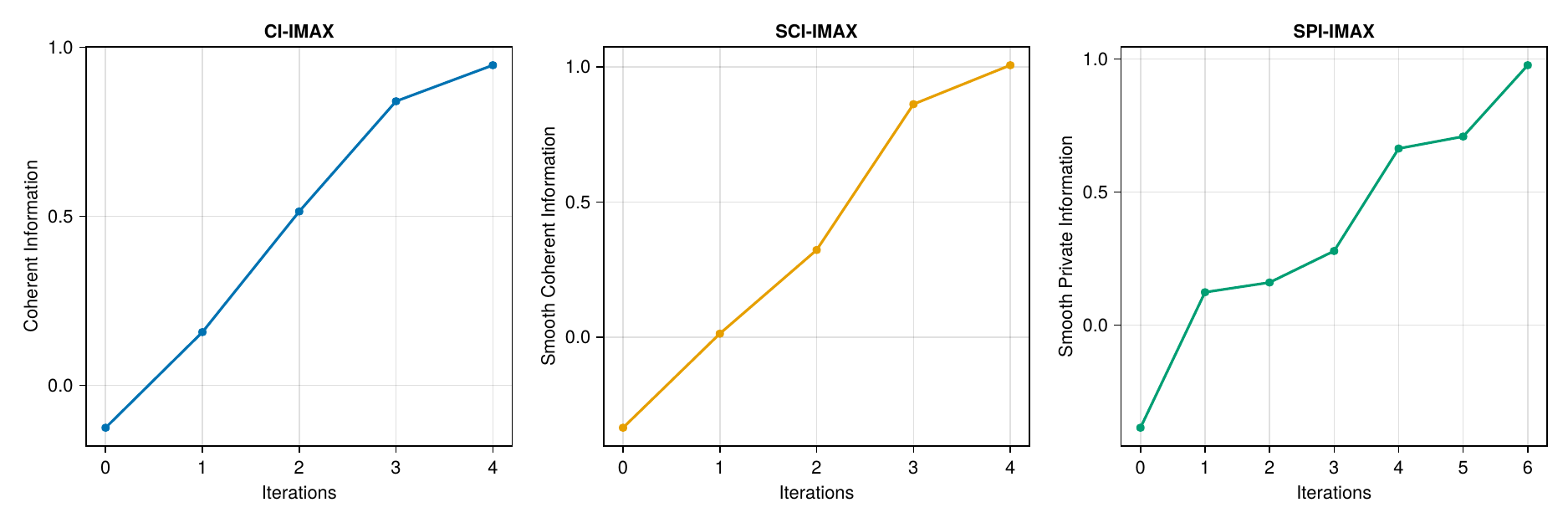}
    \caption{Resource state distillation based on three optimization quantities for $d=2$. Starting with the isotropic state $\rho(p) = p~|\Omega_{0,0}\rangle\langle \Omega_{0,0}| + (1-p)~\pi_{mm}$  for $p=0.7$ and where $\pi_{mm}$ is the maximally mixed state, different optimization quantities are iteratively increased. For the evaluation of the smooth coherent information $(-H^\varepsilon_{max})$ (used for SCI-IMAX) we use the semidefinite program (SDP) presented in \cite{nuradha_fidelity-based_2024}. For the smooth private information (used for SPI-IMAX), $I^\varepsilon_H$ can also be evaluated by an SDP \cite{khatri_principles_2024}. The smooth max-mutual information $I^{\epsilon}_{max}$ is approximated by the SDP-based algorithm given in Ref.\cite{popp_computation_2026}. The smoothing parameters $\varepsilon, \epsilon$ are both set to $0.1$. The protocols are iterated until the optimization quantity is larger than $90 \%$ of the maximally achievable value. For all protocols we use the implementation of Ref.\cite{popp_belldiagonalqudits_2023}.
    }
    \label{fig:resource_dist_example}
\end{figure}

Theorem \ref{thm:lowerboundkeyAS} states that the hypothesis testing mutual information $I^\varepsilon_H$ (Def.\ref{def:hypoEntInf}) and the smooth max-mutual information $I^\varepsilon_{max}$ (Def.\ref{def:maxrelEntInf}) define a lower bound for the one-shot distillable secret-key. More precisely, for parameters $\varepsilon$ and $\epsilon$, we consider $I^\varepsilon_H(\chi^{x,y}_{AB}) - I^{\epsilon}_{max}(\chi^{x,y}_{AE})$, where $\chi^{x,y}_{AB}$ \eqref{eq:bibartite_locdephased_AB} and $\chi^{x,y}_{AE}$ \eqref{eq:bibartite_locdephased_AE} are the dephased marginals of the tripartite stabilizer channel output. We also call this expression ``smooth private information''. As shown in Ref.\cite{popp_local_2025}, $I^\varepsilon_H$ is invariant under local isometric transformations, while $I^{\epsilon}_{max}$ is invariant under unitary transformations in the first subsystem and isometric transformations in the second system. This implies, on the one hand, that the choice of stabilizer encoding is not relevant for maximizing these quantities. On the other hand, as shown in Sec.\ref{sec:simplify_purification}, when evaluating $I^{\epsilon}_{max}$, we may consider the state $\zeta^{x,y}_{AE'}$ as in \eqref{eq:equivalent_dephased_reduced_states_AE} instead of the higher dimensional $\chi^{x,y}_{AE}$ of \eqref{eq:bibartite_locdephased_AE}. We define the ``Smooth Private Information Increase Maximizing (SPI-IMAX)'' protocol as follows.\\ \ \\
\textbf{SPI-IMAX protocol:}\\
Let $\rho_{AB}$ be any bipartite state shared between Alice and Bob with prime local dimensions $d_A = d_B$ and $\varepsilon, \epsilon \in (0,1)$.
\begin{enumerate}
    \item For each stabilizer $S$, Alice and Bob fix an arbitrary encoding $U_S$.
    \item For all $S$ and all measurement outcomes $x,y$, Alice calculates $I^{\varepsilon}_H(\zeta^{x,y}_{AB}) - I^{\epsilon}_{max}(\zeta^{x,y}_{AE})$ for the corresponding dephased tripartite post-measurement channel outputs $\zeta^{x,y}_{AB}$ and $\zeta^{x,y}_{AE}$ as in \eqref{eq:equivalent_dephased_reduced_states} and \eqref{eq:equivalent_dephased_reduced_states_AE}.
    \item Alice determines the configuration $(S_{max}, x_{max}, y_{max})$ that maximizes $I^{\varepsilon}_H(\zeta^{x,y}_{AB}) - I^{\epsilon}_{max}(\zeta^{x,y}_{AE})$ and communicates it to Bob.
    \item Alice and Bob apply the stabilizer channel corresponding to $S_{max}$ with the arbitrary encoding $U_{S_{max}}$ and determine their outcomes $x,y$. If $x \neq x_{max}$ or $y \neq y_{max}$, they declare failure of the protocol. 
\end{enumerate}
Figure \ref{fig:resource_dist_example} below includes this protocol for an isotropic input state. 

\noindent
\textbf{PI-IMAX protocol:} \\
The private information $I_p$ (see comment below Theorem \ref{thm:lowerboundkeyAS}) provides a lower bound of the distillable secret-key in the asymptotic setting. It is straightforward to define a similar protocol to the ones introduced above that maximizes the increase of private information among all stabilizer-based protocols. However, it turns out that such a protocol is equivalent to the bipartite CI-IMAX protocol that maximized the coherent information.
To see this, we again use the results of Sec. \ref{sec:simplify}. First, we have $I_p(\chi^{x,y}_{ABE}) = I_p(\zeta^{x,y}_{ABE'}) $ with the states of \eqref{eq:equivalent_dephased_state}-\eqref{eq:equivalent_dephased_reduced_states_AE}. Considering that $\zeta^{x,y}_{ABE'} := \mathcal{D}_A(\psi^{x,y}_{ABE'})$ for the purification  $\psi^{x,y}_{ABE'}$ of the bipartite output state $\sigma^{x,y}_{AB}$ , the private information of the state satisfies $I_p(\zeta^{x,y}_{ABE'}) = I^{A\rangle B}_c(\sigma^{x,y}_{AB})$ (cf. Ref.\cite{wilde_quantum_2013}). Consequently, applying the bipartite protocol CI-IMAX of \ref{sec:specific_ent_protocols} also maximizes the increase of private information, and PI-IMAX is equivalent to CI-IMAX.

\section{Conclusion}
\label{sec:conclusion}
In this work, we have formulated the stabilizer distillation routine—originally introduced in Ref.\cite{matsumoto_conversion_2003} and further developed in Ref.\cite{popp_novel_2025}—as a quantum channel. This perspective provides a unified framework for stabilizer-based methods, extending their application from bipartite entanglement distillation to tripartite secret-key distillation and general input states.\\
We derived explicit Kraus representations for both bipartite and tripartite stabilizer channels, parameterized by the choice of stabilizer and encoding. Leveraging the shared group structure of stabilizers and Bell states \cite{popp_novel_2025}, we obtained closed-form expressions for the resulting output states. These expressions establish a general framework for stabilizer-based resource distillation: for any information-processing task characterized by a specific measure, one can evaluate that measure on the channel outputs and optimize the stabilizer channel to maximize the resource increase under iterative application. Key examples demonstrated here include the (smooth) coherent information for (one-shot) entanglement distillation and the (smooth) private information for (one-shot) secret-key distillation.\\
To ensure this optimization remains computationally tractable, we leveraged local invariances within the relevant information measures. Specifically, we demonstrated invariance under changes of stabilizer encodings and, in the tripartite setting, under the choice of the purification system. These invariances significantly reduce the numerical complexity of the optimization, effectively transforming otherwise intractable tasks into feasible numerical optimizations.\\
We provided several proofs of concept illustrating the versatility of stabilizer channels for resource-state distillation. First, we generalized the FIMAX protocol \cite{popp_novel_2025} to gF-IMAX, which optimizes fidelity for arbitrary input states. Initial numerical tests on random pure states indicate that gF-IMAX can outperform the standard FIMAX protocol in specific regimes. Second, we introduced CI-IMAX and SCI-IMAX to optimize coherent information in the asymptotic and one-shot regimes, respectively. As coherent information lower-bounds the distillable entanglement, these protocols yield stabilizer channels that efficiently generate highly entangled states. Finally, we extended the framework to tripartite secret-key distillation. Noting that maximizing coherent information via CI-IMAX also maximizes the asymptotic private information, we developed the SPI-IMAX protocol to increase the smooth private information. Since this quantity provides lower bounds for one-shot secret-key rates, SPI-IMAX generates states highly optimized for one-shot cryptographic tasks. Numerical results for isotropic input states confirm that these protocols successfully identify effective operations to tailor resourceful states to specific quantum processing tasks.\\
Several directions for future research naturally follow. A systematic numerical comparison of the proposed protocols, particularly for highly mixed states, would clarify their relative efficiency and distillation effectiveness. Furthermore, identifying bipartite NPT states that are undistillable via stabilizer-based protocols yet still possess private capacity could provide new insights into the distillability problem and the existence of NPT bound entanglement. Additionally, stabilizer-based methods may help uncover PPT states with nonzero distillable secret-key, which would be instrumental in understanding bound entanglement \cite{hiesmayr_bipartite_2025} and the broader separability problem.\\
In summary, these contributions establish stabilizer channels as a versatile and optimizable tool for quantum resource distillation. By connecting stabilizer operations to the optimization of general quantum resources, this framework provides practical, high-performance protocols for both entanglement and secret-key distillation in one-shot and asymptotic regimes.
\section{Methods} \label{sec:methods}
In this section, we summarize the methods applied to generate the results presented in Sec.\ref{sec:results}.
\subsection{Stabilizer Distillation Routine}
\label{sec:stabBasedDistillation}
First, we present the background of the stabilizer framework developed in Ref.\cite{popp_novel_2025}.\\
Let $\Zd \equiv \mathbbm{Z}/d\mathbbm{Z}$ be the quotient ring of integers with addition and multiplication modulo $d$, which is a field for prime $d$ and let $\w \equiv \exp(\frac{2 \pi i}{d})$. Complex conjugation $(\star)$ is defined in the computational basis.
\begin{mydef}[Weyl(-Heisenberg) operators, Weyl(-Heisenberg) errors]\label{def:weylOpsErrs}
    \begin{flalign}
        W_{k,l} &:= \sum_{j=0}^{d-1}\w^{j k} |j\rangle \langle j+l|,~~k,l \in \Zd \\
        \E_N &:= \left\{  W(e) ~|~ e = (\Vec{k}, \Vec{l}) \in \Zd^N \times \Zd^N \right\} := \left\{ \bigotimes_{n = 1}^N W_{k_n,l_n}  ~|~ k_n,l_n \in \Zd \right\}
    \end{flalign}
\end{mydef}
\noindent 
The Weyl-Heisenberg operators satisfy the Weyl relations, i.e.,
\begin{flalign}
    \label{eq:weylRelations}
    \begin{aligned}
           W_{k_1,l_1}W_{k_2,l_2} &= \w^{l_1 k_2}~W_{k_1+k_2, l_1+l_2},  \\
           W_{k,l}^\dagger &= \w^{k l}~W_{-k, -l} = W_{k,l}^{-1},
    \end{aligned}
\end{flalign}
implying that the set of Weyl-errors $\E_N$ forms a group under multiplication, if we identify errors that are equal up to a phase, i.e., $W_{k_1,l_1}W_{k_2,l_2} \equiv W_{k_1+k_2, l_1+l_2}$. $E \in \E_N$ are called error operators. For $E = W(e)$, $e \equiv (\Vec{k}, \Vec{l}) \in \Zd^N \times \Zd^N$ are called error elements. The group structure of $\E_N$ induces a group structure via the Weyl relations \eqref{eq:weylRelations} for error elements on $\Zd^N \times \Zd^N$ with addition modulo $d$.
\begin{mydef}[Bell states]
\label{def:bell_states}
\begin{flalign}
  \label{bellStates}
     |\Okl\rangle &:= (W_{k,l} \otimes \mathbbm{1}_d)|\Omega_{0,0}\rangle := (W_{k,l} \otimes \mathbbm{1}_d)\frac{1}{\sqrt{d}}\sum_{i} | i \rangle \otimes |i \rangle,~~i,k,l \in \Zd, \\
    \Oe &:= (W(e)\otimes \id_d^{\otimes N}) |\Omega_{0,0}\rangle^{\otimes N} = \bigotimes_{n=1}^N |\Omega_{k_n,l_n} \rangle, ~~e \equiv (\Vec{k}, \Vec{l}) \in \Zd^N \times \Zd^N.
\end{flalign}
\end{mydef}
\begin{mydef}[Stabilizer group, generating operators, generating elements] \ \newline
    A stabilizer $S$  is an abelian subgroup of $\E_N$. If $\lbrace W(g_1), \dots, W(g_p) \rbrace$ forms a minimal generating set of $S$, each $W(g_j)$ is called a generating operator and we write $S = \langle W(g_1), \dots, W(g_p) \rangle$. Each $g_j \in 
    \Zd^N \times \Zd^N$ is called a generating element. Given one choice of generating elements, the corresponding subgroup of $\Zd^N \times \Z_d^N$, i.e., $G_S := \lbrace g \in \Zd^N \times \Zd^N ~|~ W(g) \in S \rbrace$ is also denoted as $G_S = \langle g_1, \dots, g_p \rangle$.
\end{mydef}
\noindent
Let $S$ be a stabilizer with generating elements $\lbrace g_1,\dots,g_p \rbrace$. For prime $d$, each generator $W(g) \neq \mathbbm{1}_d^{\otimes N}$ has $d$ distinct eigenvalues with equal multiplicity \cite{popp_novel_2025}, allowing us to associate each of the $d$ eigenspaces of a generator $W(g_j)$ with a number $x_j\in \Zd$. We write $\mQ(x)$ for the joint eigenspace of $\lbrace W(g_j) \rbrace_{j=1}^p$, such that for all $|\phi\rangle \in \mQ(x)$ and $j \in \lbrace 1, \dots, p\rbrace$, $x=(x_1,\dots,x_p) \in \Zd^p$ and $\mQ(x)\subset \hs_A^{\otimes N}$, we have $W(g_j)|\phi\rangle = \w_{x_j}|\phi\rangle$ with $x_j \in \Zd$ and $\w_{x_j}\in \mathbbm{C}$. \\
Decomposing the Hilbert space into eigenspaces of the generators,  $\hs_A^{\otimes N} \cong \bigoplus_{x\in \Zd^p} \mQ(x)$, defines the codespaces $\mQ(x)$ in $\hs_A^{\otimes N}$. Given $p \leq N$ generators of a stabilizer in $\E_N$ for prime dimensions $d$, the codespaces have dimension $\dim(\mQ(x)) = d^{N-p}~ \forall x \in \Zd^p$ \cite{popp_novel_2025}. 
Let $S^{\star}$ be the stabilizer with complex conjugated elements of $S$ and generating operators $W^{\star}(g_j)$. $\mQ^{\star}(x)\subset \hs_B^{\otimes N}$ is the joint eigenspace of $\lbrace W^{\star}(g_j) \rbrace_{j=1}^p$, such that for all $|\phi\rangle \in \mQ^{\star}(x)$ and $j \in \lbrace 1, \dots, p\rbrace$, we have $W^{\star}(g_j)|\phi\rangle =\w^\star_{x_j}|\phi\rangle$.
\begin{mydef}[Error coset] \label{def:error_coset}\ \\
Let $S \subset \E_N$ be a stabilizer and let $G_S \subset \Zd^N \times \Zd^N$ be the corresponding subgroup of error elements, such that $S = \lbrace W(g) ~|~g \in G_S \rbrace$. 
Given an error element $e \in \Zd^N \times \Zd^N$, the error coset is defined as $C(e):= e + G_S = \lbrace e + h ~|~ h\in G_S \rbrace$.
\end{mydef}
\begin{mydef}[Codeword] \label{def:codeword} \ \\
    Let $\hs_A^{\otimes N} \cong \bigoplus_{x\in \Zd^p} \mQ(x)$ be decomposed into $d^{N-p}$-dimensional codespaces of a stabilizer $S$. 
    Let $\lbrace |u_{x,k} \rangle \rbrace_{x \in \Zd^p, ~k \in \Zd^{N-p}}$ be an orthonormal basis of $\hs_A^{\otimes N}$ with $|u_{x,k} \rangle \in \mQ(x)$, i.e., $W(g_j)|u_{x,k} \rangle = w_{x_j}|u_{x,k}\rangle, ~ \forall k,j$. The vectors $| u_{x,k} \rangle \in \hs_A^{\otimes N}$ are called codewords of $S$ in $\hs_A^{\otimes N}$. The codewords of the stabilizer with complex conjugated elements $S^{\star}$ in $\hs_B^{\otimes N}$ are denoted by $| u^{\star}_{x,k} \rangle \in \hs_B^{\otimes N}$.
\end{mydef}
\begin{mydef}[Encoding] \label{def:encoding} \ \\
     Let $\lbrace |x \rangle \rbrace_{x \in \Zd^p}$ be the computational basis of $\hs_A^{\otimes p}$ and $\lbrace |k \rangle \rbrace_{k \in \Zd^{N-p}}$ be the computational basis of $\hs_A^{\otimes N-p}$. 
    A unitary operator $U$ on $\hs_A^{\otimes N} \cong \hs_A^{\otimes p} \otimes \hs_A^{\otimes N-p}$ is an encoding for a stabilizer $S$ in $\hs_A^{\otimes N}$ if $~\forall x \in \Zd^p,k\in \Zd^{N-p}:~ U (|x\rangle \otimes |k\rangle)$ is a codeword of $S$ in $\hs_A^{\otimes N}$.
\end{mydef} 
\noindent
If $U:|x\rangle \otimes |k\rangle \mapsto |u_{x,k}\rangle \in \hs_A^{\otimes N}$ is an encoding for $S$ in $\hs_A^{\otimes N}$, then $U^{\star}:|x\rangle \otimes |k\rangle \mapsto |u^{\star}_{x,k}\rangle \in \hs_B^{\otimes N}$ is an encoding for $S^{\star}$ in $\hs_B^{\otimes N}$.
\begin{mydef}[Stabilizer measurement] \label{def:stabMeasurement} \ \\
Let $\mP_A(x): \hs_A^{\otimes N} \rightarrow \mQ(x)$ be the projection to the codespace $\mQ(x)$ and $\mP_B^{\star}(y): \hs_B^{\otimes N} \rightarrow \mQ^{\star}(y)$ be the projection to the codespace $\mQ^{\star}(y)$.
The measurement corresponding to the PVM $\lbrace \mP(x,y) := \mP_A(x)\otimes \mP_B^{\star}(y) ~|~ x,y \in \Zd \rbrace$ is called stabilizer measurement.
\end{mydef}
\begin{mydef}[Symplectic product decomposition] \label{def:symprod} \ \\    
    The symplectic product of two error elements, $e=(\Vec{k},\Vec{l})$ and $f=(\Vec{m}, \Vec{n})$ is defined as $\langle e,f \rangle := \sum_{i = 0}^{N-1} l_i m_i - k_i n_i$.
    It induces a decomposition of the set of errors according to its values $s=(s_1, \dots , s_p)$ with respect to a stabilizer with generating elements $\lbrace g_1, \dots, g_p\rbrace: \E_N \cong \bigoplus_s \E(s)$, with 
    \begin{flalign}
        \E(s) := \lbrace e \in \Zd^N \times \Zd^N ~|~ \langle  g_j, e \rangle = s_j ~\forall j=1,\dots,p \rbrace.
    \end{flalign}
\end{mydef} 
\noindent
Ref.~\cite{popp_novel_2025} demonstrates that the properties of the stabilizer encodings $U$ and $U^\star$ can be leveraged to show that the state after steps 1 - 3 can be written in terms of certain unitary operators, the so-called error action operators $T^{U,e}_x$ for the error $e$, the encoding $U$, and the local stabilizer measurement outcome $x$:
\begin{lemma}[\cite{popp_novel_2025}] \label{thm:encoding_errors}
Let $U$ be an encoding of a stabilizer $S$ with generating elements $\lbrace g_1,\dots,g_p\rbrace$. For each codespace $\mQ(x) \subset \hs_A^{\otimes N}$  and for each $W(e) \in \E_N$ with $(\langle g_1, e \rangle,\dots,\langle g_p, e \rangle) = (s_1,\dots,s_p)$, there exist unitary ``action'' operators $T^{U,e}_{x}: \hs_A^{\otimes N-p}\rightarrow \hs_A^{\otimes N-p}$ satisfying
    $U^\dagger W(e) U = \sum_{x\in \Zd^p} |x+s \rangle \langle x | \otimes T_{x+s}^{U,e}$.
\end{lemma}
\begin{prop}[\cite{popp_novel_2025}]
\label{thm:standard_form_general}
    Let $S$ be a stabilizer with $p$ generating elements defining the symplectic partition of errors in $\E_N \cong \bigoplus_s \E(s)$ and $U$ be an encoding. 
    Let $\rho_{AB} = \sum_{e,f \in \E_N} \rho(e, f) ~|\Omega(e)\rangle \langle \Omega(f)|$ be a general input state in the Bell basis.
    After performing the stabilizer measurements with results $x,y$, applying $U^{-1}\otimes(U^\star)^{-1}$ and discarding the first $p$ copies, the output state $\sigma^{x,y}_{AB} \in \bigotimes_{n=p+1}^N \hs_A \otimes \hs_B$ is
    \begin{flalign}
    \label{eq:standard_form_general}
    \sigma^{x,y}_{AB} = \frac{1}{\Prob(x,y)}\sum_{e,f\in \E(x-y)}  \rho(e,f) ~(
        T_{x}^{U,e} \otimes \id)~ 
        \Pnull^{\otimes N-p}
        ~(T_{x}^{U,f} \otimes \id)^{\dagger}.
    \end{flalign}
\end{prop}

\subsection{Entanglement and Secret-Key Distillation tasks} \label{sec:distTasks}
In this section, we briefly review the tasks of entanglement distillation and secret-key distillation and define the distillable entanglement and distillable key in the asymptotic and one-shot setting.
\subsubsection{Entanglement distillation}
\label{sec:ed_task}
The goal of entanglement distillation is to (approximately) create maximally entangled states, which we define as follows:
\begin{mydef}[Maximally entangled state of Schmidt rank $D$]  \ \\
    The state 
    \begin{flalign*}
        \Psi_{AB} := \frac{1}{D} \sum_{i,j = 0}^{D-1} |i\rangle \langle j|_{A} \otimes |i\rangle \langle j|_{B}
    \end{flalign*}
    is called the maximally entangled state of Schmidt rank $D$.
\end{mydef}
\noindent Using the shared bipartite state $\rho_{AB}$, Alice and Bob apply a protocol $(D, \mathcal{L}_{AB})$, where the Schmidt-rank $D$ is the local dimension of the maximally entangled target state and $\mathcal{L}_{AB}$ is a LOCC channel as in \eqref{eq:LOCCchannel}, which they apply to $\rho_{AB}$. The entanglement distillation error is given by
\begin{flalign}
    \label{eq:entdisterror}
    p_{\mathcal{E}}^{err}(\mL_{AB}, \rho_{AB}) := 1-\mathcal{F}(\Psi_{AB}, \mL_{AB}(\rho_{AB})),
\end{flalign}
where $\Psi_{AB}$ is the maximally entangled state and $\mathcal{F}(\rho, \sigma) := \left(\Tr[\sqrt{\sqrt{\sigma} \rho \sqrt{\sigma}}]\right)^2$ is the fidelity of two mixed states \cite{uhlmann_transition_1976}. \\
The distillable entanglement of a quantum state measures how useful it is for entanglement distillation. Contrary to the asymptotic setting, in which infinitely many copies of the input state can be used for the distillation task, in the one-shot setting the LOCC channel corresponding to the protocol can only be used once. In this setting the output of the channel will generally differ by some error from the  maximally entangled target state.
\begin{mydef}[One-shot distillable entanglement] \ \\
\label{def:osDistEnt}
The one-shot distillable entanglement with distillation error $\varepsilon \in [0,1]$ of a bipartite state $\rho_{AB}$ is
    \begin{flalign*}
        \E^\varepsilon_D(\rho_{AB}) := \sup\limits_{D, \mathcal{L}_{AB}} \lbrace \log_2 D
       ~:~p_{\mathcal{E}}^{err}(\mL_{AB}, \rho_{AB}) \leq \varepsilon \rbrace,
    \end{flalign*}
    where $(D,\mathcal{L}_{AB})$ is a protocol creating approximately maximally entangled states of size $D$ using the LOCC channel $\mathcal{L}_{AB}$.
\end{mydef}
\noindent In the asymptotic setting, Alice and Bob are allowed to use an infinite number of copies of $\rho_{AB}$ to create maximally entangled states. If a protocol uses a finite number $n$ of copies of $\rho_{AB}$, it is equivalent to a protocol using the input state $\rho_{AB}^{\otimes n}$ once. This allows defining the distillable entanglement in the asymptotic setting as the maximum of so-called achievable rates for the creation of target states:
\begin{mydef}[Rate, distillable entanglement of a quantum state] \ \\
Given a protocol $(n,D, \mL_{AB})$ for a LOCC channel $\mL_{AB}$ acting on $n$ copies of a state, having $p_{\mathcal{E}}^{err}(\mL_{AB}, \rho_{AB}^{\otimes n}) \leq \varepsilon$, the rate of the protocol is $R(n,D) := \frac{\log_2 D}{n}$.
Given a state $\rho_{AB}$, a rate is achievable, if for all $\varepsilon \in (0,1], \delta > 0$ and sufficiently large $n$, there exists a protocol $(n, 2^{n(R-\delta)}, \mL_{AB})$ with $p_{\mathcal{E}}^{err}(\mL_{AB}, \rho_{AB}^{\otimes n}) \leq \varepsilon$. The quantity
\begin{flalign*}
    \mathcal{E}_D(\rho_{AB}) := \sup \lbrace R~:~R \text{ is an achievable rate for } \rho_{AB} \rbrace
\end{flalign*}
is called the distillable entanglement of $\rho_{AB}$.
\end{mydef}
\subsubsection{Secret-key distillation}
\label{sec:skd_task}
The goal of secret-key distillation is to (approximately) create tripartite key states. By definition, those states are a resource to establish a cryptographic key between two parties, which is secret from any malicious third party. For this, consider the local measurement channel corresponding to any local system $F$ or $G$ and composite system $FG$, respectively:
\begin{flalign}
    \label{eq:measurement_channels}
    &\mathcal{D}_F(\cdot) := \sum_{i} | i \rangle \langle i |_F~(\cdot)~| i \rangle \langle i |_F,
    ~~~\mathcal{D}_{FG}(\cdot) := \mathcal{D}_F(\cdot) \otimes \mathcal{D}_G(\cdot).
\end{flalign}
\begin{mydef}[Secret-key state] \ \\
    A state $\gamma_{ABE}$ is a tripartite secret-key state of size $K$ if, after application of the measurement channel 
    $\mathcal{D}_{AB}$, the state becomes a maximally correlated state of size $K$ in the systems $A$ and $B$,
\begin{flalign*}
    \hat{\Psi}_{AB} = \frac{1}{K}\sum_{i=0}^{K-1} |i\rangle\langle i|_A \otimes | i \rangle \langle i|_B,
\end{flalign*}
that is product with the state $\sigma_E$ of the system $E$:
\begin{flalign*}
    \mathcal{D}_{AB}(\gamma_{ABE})\equiv \mathcal{D}_{AB}\otimes \id_E(\gamma_{ABE}) = \hat{\Psi}_{AB} \otimes \sigma_E.
\end{flalign*}
\end{mydef}     
\noindent To create such states from a given bipartite state $\rho_{AB}$, Alice and Bob use a protocol $(K,\mL_{ABE})$, where $K$ is the size of the key system and $\mL_{ABE}$ is a LOPC channel as in \eqref{eq:LOPCchannel}, which they apply to a purification $\phi_{ABE}$ of $\rho_{AB}$. The key distillation error is given by 
\begin{flalign}
    \label{eq:disterror}
    p_{\mathcal{K}}^{err}(\mL_{ABE}, \rho_{AB}) := \inf_{\gamma_{ABE}} \left\{ 1-\mathcal{F}(\gamma_{ABE}, \mL_{ABE}(\phi_{ABE})) \right\},
\end{flalign}
where the optimization is with respect to all tripartite key states $\gamma_{ABE}$.\\
Given a quantum state, the distillable key of that state quantifies how useful it is for secret-key distillation. We consider both the asymptotic setting of infinite resources (cf., e.g., Refs.~\cite{devetak_distillation_2005, christandl_unifying_2007}) and the resource-limited one-shot setting (cf., e.g., Refs.~\cite{tomamichel_quantum_2016, khatri_second-order_2021}). While the former allows infinitely many copies of an input state to be used simultaneously for the channel, in the one-shot setting, the LOPC channel is only used once with a fixed number of input states. In this setting, secret-key states can generally only be approximately created, so we consider protocols with a finite distillation error \eqref{eq:disterror}.
\begin{mydef}[One-shot distillable key]\ \\
\label{def:osDistKey}
The one-shot distillable key with distillation error $\varepsilon \in [0,1]$ of a bipartite state $\rho_{AB}$ is
   \begin{flalign*}
       \mathcal{K}^\varepsilon_D(\rho_{AB}) := \sup\limits_{K, \mathcal{L}_{ABE}} \lbrace \log_2 K 
       ~:~p_{\mathcal{K}}^{err}(\mL_{ABE}, \rho_{AB}) \leq \varepsilon \rbrace,
   \end{flalign*}
   where $(K,\mL_{ABE})$ is a protocol creating  approximately secret-key states of size $K$ using the LOPC channel $\mL_{ABE}$.
\end{mydef} \noindent
In the asymptotic setting, Alice and Bob may use any number of copies of  $\rho_{AB}$ to produce secret-key states. If a protocol uses $n$ copies of $\rho_{AB}$, it is equivalent to a one-shot protocol using the input state $\rho_{AB}^{\otimes n}$ and we can define accordingly:
\begin{mydef}[Rate, distillable key of a quantum state] \ \\
Given a protocol $(n,K, \mL_{ABE})$ for a LOPC channel $\mL_{ABE}$ acting on $n$ copies of a state, having $p_{\mathcal{K}}^{err}(\mL_{ABE}, \rho_{AB}^{\otimes n}) \leq \varepsilon$, the rate of the protocol is $R(n,K) := \frac{\log_2 K}{n}$.
Given a state $\rho_{AB}$, a rate is achievable, if for all $\varepsilon \in (0,1], \delta > 0$ and sufficiently large $n$, there exists a protocol $(n, 2^{n(R-\delta)}, \mL_{ABE})$ with $p_{\mathcal{K}}^{err}(\mL_{ABE}, \rho_{AB}^{\otimes n}) \leq \varepsilon$. The quantity
\begin{flalign*}
    \mathcal{K}_D(\rho_{AB}) := \sup \lbrace R~:~R \text{ is an achievable rate for } \rho_{AB} \rbrace
\end{flalign*}
is called the distillable key of $\rho_{AB}$.
\end{mydef}
\subsection{Distillation bounds based on generalized information measures}
\label{sec:subsecBounds}
Given a certain bipartite input state $\rho_{AB}$, it is generally not clear if there exists a LOCC channel that transforms it to an approximately maximally entangled state. The same holds for LOPC channels to create an approximately secret-key state. However, there exist lower bounds on the (one-shot distillable entanglement and the (one-shot) distillable secret-key. These bounds are based on information measures that can be defined via generalized divergences (see Ref.\cite{khatri_principles_2024} for a comprehensive summary). In the one-shot setting, these are ``smoothed'', i.e., the environment around a given quantum state for which the measure is to be evaluated is taken into account. Often, the sine distance \cite{khatri_principles_2024} is used to define this smoothing environment.
\begin{mydef}[Smoothing environment] \label{def:smoothinBall} \ \\
Using the sine distance $P(\rho,\sigma) := \sqrt{1-\mathcal{F}(\rho,\sigma)}$, the smoothing environment $B^{\varepsilon}(\rho)$ around a state $\rho$ is defined as
    \begin{flalign}
        \label{eq:smoothBall}
        B^\varepsilon(\rho) := \left\{ \tilde{\rho} \in \denop(\hs): P(\rho,\tilde{\rho}) \leq \varepsilon \right\} 
        = \left\{ \tilde{\rho} : \mathcal{F}(\rho, \tilde{\rho}) \geq 1-\varepsilon^2 \right\}.
    \end{flalign}
\end{mydef} \noindent
Specifically, we consider the following generalized divergences and related information quantities that can be used to bound the distillable secret-key from below. 
\begin{mydef}[Quantum relative entropy, quantum mutual information, coherent information \cite{umegaki_conditional_1962, hiai_proper_1991}] \label{def:quantRelEnt}
    \begin{flalign}
        &D(\rho||\sigma) := \begin{cases}
            \Tr[\rho(\log_2\rho - \log_2\sigma)] \text{ if supp}(\rho) \subseteq \text{supp}(\sigma) \\
            +\infty ~~~~~~~~~~~~~~~~~~~~~~~ \text{ else}
        \end{cases}\\
        \label{eq:mutualInf}
        &I(\rho_{AB}) := D(\rho_{AB}||\rho_A \otimes \rho_B) \\
        &I^{A\rangle B}_c(\rho_{AB}) := D(\rho_{AB}||\id_A \otimes \rho_B)
    \end{flalign} 
\end{mydef}
\begin{mydef}[Quantum hypothesis testing relative entropy, hypothesis testing mutual information \cite{hayashi_general_2002, buscemi_quantum_2010, wang_one-shot_2012}]
\label{def:hypoEntInf}
    \begin{flalign}
    D_H^\varepsilon(\rho||\sigma) &:= -\log_2 \left( 
        \inf_{\Lambda} \left\{
            \Tr[\Lambda \sigma] : 0 \leq \Lambda \leq \mathbbm{1}, \Tr[\Lambda \rho] \geq 1-\varepsilon
        \right\}
        \right) \\
     \label{eq:hypoTestMutualIn}    
    I^\varepsilon_H(\rho_{AB}) &:= \inf_{\sigma_B}D^\varepsilon_H(\rho_{AB}||\rho_A\otimes\sigma_B),
    \end{flalign}
    where the optimization is over all quantum states $\sigma_B$.
\end{mydef}
\begin{mydef}[Max-relative entropy, smooth max-mutual information \cite{datta_min-_2009}, smooth conditional max-entropy \cite{renner_security_2006}] 
\label{def:maxrelEntInf}
    \begin{flalign}
        \label{eq:smoothMaxMutualInf}
        D_{max}(\rho||\sigma) &:= \begin{cases}
            \log_2 \left\| \sigma^{-\frac{1}{2}} \rho \sigma^{-\frac{1}{2}}  \right\|_{\infty} \text{ if supp}(\rho) \subseteq \text{supp}(\sigma) \\
            + \infty ~~~~~~~~~~~~~~~~~~~~~~~ \text{ else}
        \end{cases} \\
        I_{max}^\varepsilon(\rho_{AB}) &:= \inf_{\tilde{\rho}_{AB} \in B^\varepsilon(\rho_{AB})} D_{max}(\tilde{\rho}_{AB} || \rho_A \otimes \tilde{\rho}_B), \\
        H^\varepsilon_{max}(\rho_{AB}) &:= \inf_{
            \substack{\tilde{\rho}_{AE} \in B^\varepsilon(\rho_{AE})\\
            \sigma_E
            }
        } D_{max}(\tilde{\rho}_{AE} || \id_A \otimes \sigma_E)
    \end{flalign}
    where $\sigma_E$ is a quantum state, $\rho_{AE} := \Tr_B[\phi_{ABE}]$ is defined via the purification $\phi_{ABE}$ of $\rho_{AB}$ and $B^\varepsilon(\rho_{AB})$ is as in Def.~\ref{def:smoothinBall}.
\end{mydef} \noindent 
Note, that the definition of $H^\varepsilon_{max}$ of Def.~\ref{def:maxrelEntInf} is based on the smooth conditional min-entropy $H^\varepsilon_{min}$ and the duality $H^\varepsilon_{max}(\rho_{AB}) = - H^{\varepsilon}_{min}(\rho_{AE})$. Consult Refs.~\cite{khatri_principles_2024} and \cite{tomamichel_quantum_2016} for more details. \\
We are now in a position to state the lower bounds on the one-shot and asymptotic distillable entanglement and distillable key,  motivating the information quantities under consideration in this work.
\begin{theorem}[Lower bound for the one-shot distillable entanglement \cite{khatri_principles_2024, wilde_converse_2017}]\label{thm:lowerboundOSEnt}\ \\
Let $\rho_{AB}$ be a bipartite state. Let $\mL$ be any LOCC channel with output $\mL(\rho_{AB}) = \hat\rho_{AB}$, $\varepsilon \in (0,1]$, and $\eta\in [0, \sqrt{\varepsilon}]$. It holds that:
\begin{flalign}
    \label{eq:lowerboundOSDistEnt}
    \mathcal{E}_D^\varepsilon(\rho_{AB}) \geq 
    \sup_{\substack{
        \mL
        }
    }
    \left(
    -H^{\sqrt{\varepsilon}-\eta}_{max}(\hat\rho_{AB}) + 4\log_2(\eta)
    \right).
\end{flalign}
\end{theorem}
\begin{theorem}[Characterization of the distillable entanglement \cite{devetak_distillation_2005, khatri_principles_2024}] \ \\
\label{thm:lowerboundentAS}
Let $\rho_{AB}$ be a bipartite state. Let $\mL^{(n)}$ be any LOCC channel acting on $n$ copies of $\rho_{AB}$ with output $\mL^{(n)}(\rho^{\otimes n}_{AB}) = \hat\rho_{AB}$. It holds that:
\begin{flalign}
    \mathcal{E}_D(\rho_{AB}) = \lim_{n\rightarrow \infty} \frac{1}{n} \sup_{\mL^{(n)}}
    I^{A \rangle B}_c(\hat\rho_{AB}).
\end{flalign}
\end{theorem}
\noindent 
For secret-key distillation, the bounds are based on so-called classical-quantum-quantum states, having the form $\chi_{ABE} = \sum_i |i\rangle\langle i |_A \otimes \chi_{BE}^i$.
\begin{theorem}[Lower bound for the one-shot distillable key \cite{khatri_principles_2024, khatri_second-order_2021}]\label{thm:lowerboundOSS}\ \\
Let $\rho_{AB}$ be a bipartite state with purification $\phi_{ABE}$. Let $\varepsilon \in (0,1]$, $\mL$ be any LOPC channel with classical-quantum-quantum output $\mL(\phi_{ABE}) = \chi_{ABE}$ and $\varepsilon' = 1 - \sqrt{1-\varepsilon}$. It holds that:
\begin{flalign}
    \label{eq:lowerboundkey}
    \mathcal{K}_D^\varepsilon(\rho_{AB}) \geq 
    \sup_{\substack{
        \mL,\\
        \delta \in (0, \varepsilon'),\\ 
        \eta \in (0, \varepsilon'-\delta),\\
        \zeta \in (0, \delta)
        }
    }
    \left(
    I_H^{\varepsilon'-\delta-\eta}(\chi_{AB}) -  I_{max}^{\delta-\zeta}(\chi_{AE}) - \log_2(\frac{4(\varepsilon'-\delta)}{\eta^2}) - \log_2(\frac{2}{\zeta^2})
    \right).
\end{flalign}
\end{theorem}
\begin{theorem}[Characterization of the distillable key \cite{devetak_distillation_2005, khatri_principles_2024}]\label{thm:lowerboundkeyAS} \ \\
Let $\rho_{AB}$ be a bipartite state with purification $\phi_{ABE}$. Let $\mL^{(n)}$ be any LOPC channel acting on $n$ copies of $\phi_{ABE}$ with classical-quantum-quantum output $\mL^{(n)}(\phi^{\otimes n}_{ABE}) = \chi_{ABE}$. It holds that:
\begin{flalign}
    \mathcal{K}_D(\rho_{AB}) = \lim_{n\rightarrow \infty} \frac{1}{n} \sup_{\mL^{(n)}}
    \left(
    I(\chi_{AB}) - I(\chi_{AE})
    \right).
\end{flalign}
\end{theorem}
\noindent The quantity $I_p(\chi_{ABE}) := I(\chi_{AB}) - I(\chi_{AE})$ is also called private information.
\printbibliography

@article{popp_computation_2026,
	title = {Computation of the smooth max-mutual information via semidefinite programming},
	volume = {25},
	issn = {1573-1332},
	url = {https://doi.org/10.1007/s11128-026-05101-8},
	doi = {10.1007/s11128-026-05101-8},
	language = {en},
	number = {3},
	urldate = {2026-02-26},
	journal = {Quantum Information Processing},
	author = {Popp, Christopher and Sutter, Tobias C. and Hiesmayr, Beatrix C.},
	month = feb,
	year = {2026},
	pages = {86},
}

@article{divincenzo_locking_2004,
	title = {Locking {Classical} {Correlations} in {Quantum} {States}},
	volume = {92},
	url = {https://link.aps.org/doi/10.1103/PhysRevLett.92.067902},
	doi = {10.1103/PhysRevLett.92.067902},
	number = {6},
	urldate = {2026-02-12},
	journal = {Physical Review Letters},
	author = {DiVincenzo, David P. and Horodecki, Michał and Leung, Debbie W. and Smolin, John A. and Terhal, Barbara M.},
	month = feb,
	year = {2004},
	note = {Publisher: American Physical Society},
	pages = {067902},
}

@article{popp_novel_2025,
	title = {A {Novel} {Stabilizer}-based {Entanglement} {Distillation} {Protocol} for {Qudits}},
	volume = {9},
	url = {https://quantum-journal.org/papers/q-2025-12-15-1945/},
	doi = {10.22331/q-2025-12-15-1945},
	language = {en-GB},
	urldate = {2025-12-18},
	journal = {Quantum},
	author = {Popp, Christopher and Sutter, Tobias C. and Hiesmayr, Beatrix C.},
	month = dec,
	year = {2025},
	note = {Publisher: Verein zur Förderung des Open Access Publizierens in den Quantenwissenschaften},
	pages = {1945},
}

@article{alber_efficient_2001,
	title = {Efficient bipartite quantum state purification in arbitrary dimensional {Hilbert} spaces},
	volume = {34},
	issn = {0305-4470},
	url = {https://doi.org/10.1088/0305-4470/34/42/307},
	doi = {10.1088/0305-4470/34/42/307},
	language = {en},
	number = {42},
	urldate = {2025-12-10},
	journal = {Journal of Physics A: Mathematical and General},
	author = {Alber, Gernot and Delgado, Aldo and Gisin, Nicolas and Jex, Igor},
	month = oct,
	year = {2001},
	pages = {8821},
}

@article{bennett_quantum_2014-1,
	series = {Theoretical {Aspects} of {Quantum} {Cryptography} – celebrating 30 years of {BB84}},
	title = {Quantum cryptography: {Public} key distribution and coin tossing},
	volume = {560},
	issn = {0304-3975},
	shorttitle = {Quantum cryptography},
	url = {https://www.sciencedirect.com/science/article/pii/S0304397514004241},
	doi = {10.1016/j.tcs.2014.05.025},
	urldate = {2025-11-04},
	journal = {Theoretical Computer Science},
	author = {Bennett, Charles H. and Brassard, Gilles},
	month = dec,
	year = {2014},
	pages = {7--11},
}

@article{popp_low-fidelity_2025,
	title = {Low-fidelity entanglement distillation with {FIMAX}},
	volume = {23},
	issn = {0219-7499},
	url = {https://www.worldscientific.com/doi/10.1142/S0219749925500170},
	doi = {10.1142/S0219749925500170},
	number = {06},
	urldate = {2025-10-24},
	journal = {International Journal of Quantum Information},
	author = {Popp, Christopher and Sutter, Tobias C. and Hiesmayr, Beatrix C.},
	month = sep,
	year = {2025},
	note = {Publisher: World Scientific Publishing Co.},
	pages = {2550017},
}

@article{hiesmayr_bipartite_2025,
	title = {Bipartite bound entanglement},
	volume = {23},
	issn = {0219-7499},
	url = {https://www.worldscientific.com/doi/10.1142/S0219749925300037},
	doi = {10.1142/S0219749925300037},
	number = {05},
	urldate = {2025-10-23},
	journal = {International Journal of Quantum Information},
	author = {Hiesmayr, Beatrix C. and Popp, Christopher and Sutter, Tobias C.},
	month = aug,
	year = {2025},
	note = {Publisher: World Scientific Publishing Co.},
	pages = {2530003},
}

@article{ashikhmin_nonbinary_2001,
	title = {Nonbinary quantum stabilizer codes},
	volume = {47},
	issn = {1557-9654},
	url = {https://ieeexplore.ieee.org/document/959288},
	doi = {10.1109/18.959288},
	number = {7},
	urldate = {2025-10-22},
	journal = {IEEE Transactions on Information Theory},
	author = {Ashikhmin, A. and Knill, E.},
	month = nov,
	year = {2001},
	pages = {3065--3072},
}

@article{popp_local_2025,
	title = {Local {Invariance} of {Divergence}-{Based} {Quantum} {Information} {Measures}},
	volume = {27},
	copyright = {http://creativecommons.org/licenses/by/3.0/},
	issn = {1099-4300},
	url = {https://www.mdpi.com/1099-4300/27/10/1051},
	doi = {10.3390/e27101051},
	language = {en},
	number = {10},
	urldate = {2025-10-20},
	journal = {Entropy},
	author = {Popp, Christopher and Sutter, Tobias C. and Hiesmayr, Beatrix C.},
	month = oct,
	year = {2025},
	note = {Publisher: Multidisciplinary Digital Publishing Institute},
	pages = {1051},
}

@article{tavakoli_semidefinite_2024,
	title = {Semidefinite programming relaxations for quantum correlations},
	volume = {96},
	url = {https://link.aps.org/doi/10.1103/RevModPhys.96.045006},
	doi = {10.1103/RevModPhys.96.045006},
	number = {4},
	urldate = {2025-09-03},
	journal = {Reviews of Modern Physics},
	author = {Tavakoli, Armin and Pozas-Kerstjens, Alejandro and Brown, Peter and Araújo, Mateus},
	month = dec,
	year = {2024},
	note = {Publisher: American Physical Society},
	pages = {045006},
}

@book{skrzypczyk_semidefinite_2023,
	title = {Semidefinite {Programming} in {Quantum} {Information} {Science}},
	isbn = {978-0-7503-3343-6},
	url = {https://iopscience.iop.org/book/mono/978-0-7503-3343-6},
	language = {en},
	urldate = {2025-09-03},
	publisher = {IOP Publishing},
	author = {Skrzypczyk, Paul and Cavalcanti, Daniel},
	month = mar,
	year = {2023},
}

@article{hiai_proper_1991,
	title = {The proper formula for relative entropy and its asymptotics in quantum probability},
	volume = {143},
	issn = {1432-0916},
	url = {https://doi.org/10.1007/BF02100287},
	doi = {10.1007/BF02100287},
	language = {en},
	number = {1},
	urldate = {2025-07-30},
	journal = {Communications in Mathematical Physics},
	author = {Hiai, Fumio and Petz, Dénes},
	month = dec,
	year = {1991},
	pages = {99--114},
}

@article{umegaki_conditional_1962,
	title = {Conditional expectation in an operator algebra. {IV}. {Entropy} and information},
	volume = {14},
	issn = {0023-2599},
	url = {https://projecteuclid.org/journals/kodai-mathematical-seminar-reports/volume-14/issue-2/Conditional-expectation-in-an-operator-algebra-IV-Entropy-and-information/10.2996/kmj/1138844604.full},
	doi = {10.2996/kmj/1138844604},
	number = {2},
	urldate = {2025-07-30},
	journal = {Kodai Mathematical Seminar Reports},
	author = {Umegaki, Hisaharu},
	month = jan,
	year = {1962},
	note = {Publisher: Institute of Science Tokyo, Department of Mathematics},
	pages = {59--85},
}

@article{uhlmann_transition_1976,
	title = {The “transition probability” in the state space of a *-algebra},
	volume = {9},
	issn = {0034-4877},
	url = {https://www.sciencedirect.com/science/article/pii/0034487776900604},
	doi = {10.1016/0034-4877(76)90060-4},
	number = {2},
	urldate = {2025-07-30},
	journal = {Reports on Mathematical Physics},
	author = {Uhlmann, A.},
	month = apr,
	year = {1976},
	pages = {273--279},
}

@misc{renner_security_2006,
	title = {Security of {Quantum} {Key} {Distribution}},
	url = {http://arxiv.org/abs/quant-ph/0512258},
	doi = {10.48550/arXiv.quant-ph/0512258},
	urldate = {2025-07-24},
	publisher = {arXiv},
	author = {Renner, Renato},
	month = jan,
	year = {2006},
	note = {arXiv:quant-ph/0512258},
}

@article{ekert_quantum_1991,
	title = {Quantum cryptography based on {Bell}'s theorem},
	volume = {67},
	url = {https://link.aps.org/doi/10.1103/PhysRevLett.67.661},
	doi = {10.1103/PhysRevLett.67.661},
	number = {6},
	urldate = {2025-07-24},
	journal = {Physical Review Letters},
	author = {Ekert, Artur K.},
	month = aug,
	year = {1991},
	note = {Publisher: American Physical Society},
	pages = {661--663},
}

@inproceedings{christandl_unifying_2007,
	address = {Berlin, Heidelberg},
	title = {Unifying {Classical} and {Quantum} {Key} {Distillation}},
	isbn = {978-3-540-70936-7},
	doi = {10.1007/978-3-540-70936-7_25},
	language = {en},
	booktitle = {Theory of {Cryptography}},
	publisher = {Springer},
	author = {Christandl, Matthias and Ekert, Artur and Horodecki, Michał and Horodecki, Paweł and Oppenheim, Jonathan and Renner, Renato},
	editor = {Vadhan, Salil P.},
	year = {2007},
	pages = {456--478},
}

@book{wilde_quantum_2013,
	address = {Cambridge},
	title = {Quantum {Information} {Theory}},
	url = {https://www.cambridge.org/core/books/quantum-information-theory/9DC2CA59F45636D4F0F30D971B677623},
	urldate = {2025-07-23},
	publisher = {Cambridge University Press},
	author = {Wilde, Mark M.},
	year = {2013},
	doi = {10.1017/CBO9781139525343},
}

@article{wang_one-shot_2012,
	title = {One-{Shot} {Classical}-{Quantum} {Capacity} and {Hypothesis} {Testing}},
	volume = {108},
	url = {https://link.aps.org/doi/10.1103/PhysRevLett.108.200501},
	doi = {10.1103/PhysRevLett.108.200501},
	number = {20},
	urldate = {2025-07-23},
	journal = {Physical Review Letters},
	author = {Wang, Ligong and Renner, Renato},
	month = may,
	year = {2012},
	note = {Publisher: American Physical Society},
	pages = {200501},
}

@article{buscemi_quantum_2010,
	title = {The {Quantum} {Capacity} of {Channels} {With} {Arbitrarily} {Correlated} {Noise}},
	volume = {56},
	issn = {1557-9654},
	url = {https://ieeexplore.ieee.org/document/5429118},
	doi = {10.1109/TIT.2009.2039166},
	number = {3},
	urldate = {2025-07-23},
	journal = {IEEE Transactions on Information Theory},
	author = {Buscemi, Francesco and Datta, Nilanjana},
	month = mar,
	year = {2010},
	pages = {1447--1460},
}

@inproceedings{hayashi_general_2002,
	title = {A general formula for the classical capacity of a general quantum channel},
	url = {https://ieeexplore.ieee.org/document/1023343},
	doi = {10.1109/ISIT.2002.1023343},
	urldate = {2025-07-23},
	booktitle = {Proceedings {IEEE} {International} {Symposium} on {Information} {Theory},},
	author = {Hayashi, M. and Nagaoka, H.},
	month = jun,
	year = {2002},
	pages = {71--},
}

@article{datta_min-_2009,
	title = {Min- and {Max}-{Relative} {Entropies} and a {New} {Entanglement} {Monotone}},
	volume = {55},
	issn = {1557-9654},
	url = {https://ieeexplore.ieee.org/document/4957651},
	doi = {10.1109/TIT.2009.2018325},
	number = {6},
	urldate = {2025-07-23},
	journal = {IEEE Transactions on Information Theory},
	author = {Datta, Nilanjana},
	month = jun,
	year = {2009},
	pages = {2816--2826},
}

@misc{khatri_second-order_2021,
	title = {Second-order coding rates for key distillation in quantum key distribution},
	url = {http://arxiv.org/abs/1910.03883},
	doi = {10.48550/arXiv.1910.03883},
	urldate = {2025-07-17},
	publisher = {arXiv},
	author = {Khatri, Sumeet and Kaur, Eneet and Guha, Saikat and Wilde, Mark M.},
	month = jul,
	year = {2021},
	note = {arXiv:1910.03883 [quant-ph]},
}

@article{popp_belldiagonalqudits_2023,
	title = {{BellDiagonalQudits}: {A} package for entanglement analyses of mixed maximally entangled qudits},
	volume = {8},
	issn = {2475-9066},
	shorttitle = {{BellDiagonalQudits}},
	url = {https://joss.theoj.org/papers/10.21105/joss.04924},
	doi = {10.21105/joss.04924},
	language = {en},
	number = {81},
	urldate = {2024-02-06},
	journal = {Journal of Open Source Software},
	author = {Popp, Christopher},
	month = jan,
	year = {2023},
	pages = {4924},
}

@article{matsumoto_conversion_2003,
	title = {Conversion of a general quantum stabilizer code to an entanglement distillation protocol},
	volume = {36},
	issn = {0305-4470, 1361-6447},
	url = {http://arxiv.org/abs/quant-ph/0209091},
	doi = {10.1088/0305-4470/36/29/316},
	language = {en},
	number = {29},
	urldate = {2024-02-06},
	journal = {Journal of Physics A: Mathematical and General},
	author = {Matsumoto, Ryutaroh},
	month = jul,
	year = {2003},
	note = {arXiv:quant-ph/0209091},
	pages = {8113--8127},
}

@article{chitambar_quantum_2019,
	title = {Quantum resource theories},
	volume = {91},
	url = {https://link.aps.org/doi/10.1103/RevModPhys.91.025001},
	doi = {10.1103/RevModPhys.91.025001},
	number = {2},
	urldate = {2025-02-06},
	journal = {Reviews of Modern Physics},
	author = {Chitambar, Eric and Gour, Gilad},
	month = apr,
	year = {2019},
	note = {Publisher: American Physical Society},
	pages = {025001},
}

@article{nuradha_fidelity-based_2024,
	title = {Fidelity-{Based} {Smooth} {Min}-{Relative} {Entropy}: {Properties} and {Applications}},
	volume = {70},
	issn = {0018-9448, 1557-9654},
	shorttitle = {Fidelity-{Based} {Smooth} {Min}-{Relative} {Entropy}},
	url = {http://arxiv.org/abs/2305.05859},
	doi = {10.1109/TIT.2024.3378590},
	number = {6},
	urldate = {2024-12-18},
	journal = {IEEE Transactions on Information Theory},
	author = {Nuradha, Theshani and Wilde, Mark M.},
	month = jun,
	year = {2024},
	note = {arXiv:2305.05859 [quant-ph]},
	pages = {4170--4196},
}

@book{tomamichel_quantum_2016,
	title = {Quantum {Information} {Processing} with {Finite} {Resources} -- {Mathematical} {Foundations}},
	volume = {5},
	url = {http://arxiv.org/abs/1504.00233},
	urldate = {2024-12-05},
	author = {Tomamichel, Marco},
	year = {2016},
	doi = {10.1007/978-3-319-21891-5},
	note = {arXiv:1504.00233 [quant-ph]},
}

@misc{divincenzo_quantum_2001,
	title = {Quantum {Data} {Hiding}},
	url = {http://arxiv.org/abs/quant-ph/0103098},
	doi = {10.48550/arXiv.quant-ph/0103098},
	urldate = {2024-11-01},
	publisher = {arXiv},
	author = {DiVincenzo, David P. and Leung, Debbie W. and Terhal, Barbara M.},
	month = mar,
	year = {2001},
	note = {arXiv:quant-ph/0103098},
}

@article{wilde_converse_2017,
	title = {Converse {Bounds} for {Private} {Communication} {Over} {Quantum} {Channels}},
	volume = {63},
	issn = {1557-9654},
	url = {https://ieeexplore.ieee.org/document/7807212/?arnumber=7807212},
	doi = {10.1109/TIT.2017.2648825},
	number = {3},
	urldate = {2024-10-02},
	journal = {IEEE Transactions on Information Theory},
	author = {Wilde, Mark M. and Tomamichel, Marco and Berta, Mario},
	month = mar,
	year = {2017},
	note = {Conference Name: IEEE Transactions on Information Theory},
	pages = {1792--1817},
}

@article{horodecki_general_2009,
	title = {General {Paradigm} for {Distilling} {Classical} {Key} {From} {Quantum} {States}},
	volume = {55},
	issn = {1557-9654},
	url = {https://ieeexplore.ieee.org/document/4802308},
	doi = {10.1109/TIT.2008.2009798},
	number = {4},
	urldate = {2024-08-28},
	journal = {IEEE Transactions on Information Theory},
	author = {Horodecki, Karol and Horodecki, Michal and Horodecki, Pawel and Oppenheim, Jonathan},
	month = apr,
	year = {2009},
	note = {Conference Name: IEEE Transactions on Information Theory},
	pages = {1898--1929},
}

@misc{khatri_principles_2024,
	title = {Principles of {Quantum} {Communication} {Theory}: {A} {Modern} {Approach}},
	shorttitle = {Principles of {Quantum} {Communication} {Theory}},
	url = {http://arxiv.org/abs/2011.04672},
	doi = {10.48550/arXiv.2011.04672},
	urldate = {2024-08-19},
	publisher = {arXiv},
	author = {Khatri, Sumeet and Wilde, Mark M.},
	month = feb,
	year = {2024},
	note = {arXiv:2011.04672 [cond-mat, physics:hep-th, physics:math-ph, physics:quant-ph]},
}

@article{devetak_distillation_2005,
	title = {Distillation of secret key and entanglement from quantum states},
	volume = {461},
	issn = {1364-5021, 1471-2946},
	url = {http://arxiv.org/abs/quant-ph/0306078},
	doi = {10.1098/rspa.2004.1372},
	number = {2053},
	urldate = {2024-08-16},
	journal = {Proceedings of the Royal Society A: Mathematical, Physical and Engineering Sciences},
	author = {Devetak, Igor and Winter, Andreas},
	month = jan,
	year = {2005},
	note = {arXiv:quant-ph/0306078},
	pages = {207--235},
}

@article{rains_nonbinary_1999,
	title = {Nonbinary quantum codes},
	volume = {45},
	issn = {1557-9654},
	url = {https://ieeexplore.ieee.org/document/782103},
	doi = {10.1109/18.782103},
	number = {6},
	urldate = {2024-08-01},
	journal = {IEEE Transactions on Information Theory},
	author = {Rains, E.M.},
	month = sep,
	year = {1999},
	note = {Conference Name: IEEE Transactions on Information Theory},
	pages = {1827--1832},
}

@misc{horodecki_reduction_1998,
	title = {Reduction criterion of separability and limits for a class of protocols of entanglement distillation},
	url = {http://arxiv.org/abs/quant-ph/9708015},
	language = {en},
	urldate = {2024-07-26},
	publisher = {arXiv},
	author = {Horodecki, Michal and Horodecki, Pawel},
	month = apr,
	year = {1998},
	note = {arXiv:quant-ph/9708015},
}

@article{deutsch_quantum_1996,
	title = {Quantum {Privacy} {Amplification} and the {Security} of {Quantum} {Cryptography} over {Noisy} {Channels}},
	volume = {77},
	url = {https://link.aps.org/doi/10.1103/PhysRevLett.77.2818},
	doi = {10.1103/PhysRevLett.77.2818},
	number = {13},
	urldate = {2024-07-17},
	journal = {Physical Review Letters},
	author = {Deutsch, David and Ekert, Artur and Jozsa, Richard and Macchiavello, Chiara and Popescu, Sandu and Sanpera, Anna},
	month = sep,
	year = {1996},
	note = {Publisher: American Physical Society},
	pages = {2818--2821},
}

@article{bennett_purification_1996,
	title = {Purification of {Noisy} {Entanglement} and {Faithful} {Teleportation} via {Noisy} {Channels}},
	volume = {76},
	issn = {0031-9007, 1079-7114},
	url = {https://link.aps.org/doi/10.1103/PhysRevLett.76.722},
	doi = {10.1103/PhysRevLett.76.722},
	language = {en},
	number = {5},
	urldate = {2024-02-06},
	journal = {Physical Review Letters},
	author = {Bennett, Charles H. and Brassard, Gilles and Popescu, Sandu and Schumacher, Benjamin and Smolin, John A. and Wootters, William K.},
	month = jan,
	year = {1996},
	pages = {722--725},
}

@article{horodecki_quantum_2009,
	title = {Quantum entanglement},
	volume = {81},
	issn = {0034-6861, 1539-0756},
	url = {http://arxiv.org/abs/quant-ph/0702225},
	doi = {10.1103/RevModPhys.81.865},
	language = {en},
	number = {2},
	urldate = {2024-02-06},
	journal = {Reviews of Modern Physics},
	author = {Horodecki, Ryszard and Horodecki, Pawel and Horodecki, Michal and Horodecki, Karol},
	month = jun,
	year = {2009},
	note = {arXiv:quant-ph/0702225},
	pages = {865--942},
}

@article{vollbrecht_efficient_2003,
	title = {Efficient distillation beyond qubits},
	volume = {67},
	issn = {1050-2947, 1094-1622},
	url = {https://link.aps.org/doi/10.1103/PhysRevA.67.012303},
	doi = {10.1103/PhysRevA.67.012303},
	language = {en},
	number = {1},
	urldate = {2024-02-06},
	journal = {Physical Review A},
	author = {Vollbrecht, Karl Gerd H. and Wolf, Michael M.},
	month = jan,
	year = {2003},
	pages = {012303},
}

@article{miguel-ramiro_efficient_2018,
	title = {Efficient entanglement purification protocols for d -level systems},
	volume = {98},
	issn = {2469-9926, 2469-9934},
	url = {https://link.aps.org/doi/10.1103/PhysRevA.98.042309},
	doi = {10.1103/PhysRevA.98.042309},
	language = {en},
	number = {4},
	urldate = {2024-02-06},
	journal = {Physical Review A},
	author = {Miguel-Ramiro, J. and Dür, W.},
	month = oct,
	year = {2018},
	pages = {042309},
}

@article{dehaene_local_2003,
	title = {Local permutations of products of {Bell} states and entanglement distillation},
	volume = {67},
	issn = {1050-2947, 1094-1622},
	url = {https://link.aps.org/doi/10.1103/PhysRevA.67.022310},
	doi = {10.1103/PhysRevA.67.022310},
	language = {en},
	number = {2},
	urldate = {2024-02-06},
	journal = {Physical Review A},
	author = {Dehaene, Jeroen and Van Den Nest, Maarten and De Moor, Bart and Verstraete, Frank},
	month = feb,
	year = {2003},
	pages = {022310},
}

\section*{Author Contributions} Conceptualization, C.P.; validation, C.P., T.C.S. and  B.C.H.; formal analysis, C.P.; writing---original draft preparation, C.P.; writing---review and editing, C.P., T.C.S. and B.C.H. All authors have read and agreed to the published version of the manuscript.

\section*{Acknowledgments} This research was funded in whole, or in part, by the Austrian Science Fund (FWF) [10.55776/P36102]. For the purpose of open access, the author has applied a CC BY public copyright license to any Author Accepted Manuscript version arising from this submission.

\end{document}